\documentclass[12pt]{article}
\pdfoutput=1
\usepackage{epsfig,xcolor}
\usepackage{epstopdf}
\usepackage{amsmath}
\usepackage{latexsym}
\usepackage{textcomp}
\usepackage{psfrag}

\usepackage{amssymb}
\usepackage{amsfonts}
\usepackage{amsmath}
\usepackage[all]{xy}
\usepackage{amsfonts}
\usepackage{amssymb,amsmath}
\usepackage{hyperref}
\usepackage{natbib}

\usepackage[margin=1in]{geometry}
\usepackage[singlespacing]{setspace}

\newtheorem{definition}{\sc{Definition}}

\newtheorem{theorem}{\sc {Theorem}}
\newtheorem{lemma}{\sc {Lemma}}

\newcommand{\qed}{\hfill \mbox{\rule{2mm}{2mm}} \vspace{0.5cm}}

\newcommand{\argmax}{{\rm argmax}}

 \bibpunct[, ]{(}{)}{,}{a}{}{,}%
\usepackage{graphicx,hyperref}
\usepackage{epsfig}
\usepackage{epstopdf}
\usepackage{caption}
\usepackage{subcaption}

\newcommand{\Xomit}[1]{}

\newcommand{\highlight}{}

\begin{document}

\title{The Cost of Free Spectrum}

\author{Th\`anh Nguyen\footnote{Contacting author, Krannert School of Management, Purdue University, nguye161@purdue.edu} \\ 
Hang Zhou, Randall A. Berry, Michael L. Honig \footnote{Northwestern University} \\
Rakesh Vohra\footnote{University of Pennsylvania}}

\date{}

\maketitle
\onehalfspacing

\begin{abstract}
There has been growing interest in increasing the amount of radio spectrum 
available for unlicensed broadband wireless access. 
That includes ``prime" spectrum at lower frequencies, which is also suitable
for wide area coverage by licensed cellular providers.
While additional unlicensed spectrum would allow for
market expansion, it could influence competition among providers
and increase congestion (interference) among consumers of
wireless services. We study the value (social welfare and consumer surplus)
obtained by adding unlicensed spectrum to an existing allocation
of licensed spectrum among incumbent service providers.
We assume a population of customers who choose a provider
based on the minimum {\it delivered price}, given by the weighted sum of the 
price of the service and a congestion cost, which depends
on the number of subscribers in a band.  We consider models in which this weighting 
is uniform across the customer population and where the weighting is either
high or low, reflecting different sensitivities to latency.
For the models considered, we find that the social welfare
depends on the amount of additional unlicensed spectrum,
and can actually decrease over a significant range
of unlicensed bandwidths. Furthermore, with nonuniform weighting,
introducing unlicensed spectrum can also reduce consumer welfare.
\end{abstract}

\section{Introduction}
The increase in demand for mobile data, driven in part 
by the proliferation of smart phones and tablets, is straining
the capabilities of current broadband wireless networks.
Service providers have consequently requested increases
in the amount of spectrum allocated to commercial broadband services.
That has motivated numerous discussions concerning policies
that would facilitate more efficient use of spectrum
\cite{PCAST,Peha09,Hazlett,ComMag,BOS10}. A key policy 
distinction is whether such new spectrum is 
{\it licensed} or {\it unlicensed}. Licensed spectrum 
provides the license holder with exclusive access and is used, 
for example, to provide cellular services. 
Unlicensed spectrum (also referred to as ``open access" or ``commons" spectrum)
can be used by any device (e.g., for WiFi access) that abides by certain 
technical restrictions, such as a limit on transmit power.

The unlicensed bands currently used by WiFi devices are 
at relatively high frequencies (i.e., 2.4 GHz and 5 GHz) 
and operate with low power restrictions, which limits 
their range to relatively short distances compared 
to the wide-area coverage of cellular services.
There has been recent interest in allocating additional
unlicensed spectrum at lower frequencies, in particular, the unused channels, or ``white spaces'' that
lie within spectrum allocated to 
broadcast television.\footnote{In 2010 the United States 
Federal Communications Commission (FCC)  
published final rules for unlicensed use of white spaces (\cite{FCC10174}).}   
Because radio signals tend to propagate further at lower frequencies,
the broadcast television bands are more suitable for wide-area coverage
than the current WiFi bands.\footnote{For example, the IEEE 802.22 standard 
being developed for white spaces can support distances
greater than 30 km, similar to commercial cellular (\cite{IEEE802-22}).} 
Allocating these bands for unlicensed access would lower the
barriers faced by new entrants seeking to provide wireless data services.
This is in contrast to licensed spectrum for cellular service,
which must be purchased by auction or by negotiations with
another Service Provider (SP), posing a steep barrier to 
entry.\footnote{For example, in 2008 firms paid more 
that \$19 billion in an auction 
for 1090 licenses within the 700 Mhz band.
The majority of those licenses were purchased by 
incumbent cellular providers \cite{FCC_auction73}.}  

Adding new entrants to the market increases competition, 
leading proponents for unlicensed spectrum to argue that 
it will benefit consumers as well as the overall economy 
(e.g., see \cite{MilgromLevin}). However, spectrum is a congestible
resource in the sense that shared use generates externalities
due to interference. Hence the Quality of Service (QoS) for
a particular user (measured in terms of throughput and/or latency) 
generally degrades as the number of users sharing the 
spectrum increases.\footnote{In addition, the network infrastructure, 
which must carry the wireless traffic, also has a capacity limit, 
potentially introducing another source of congestion.
Here we are mainly concerned with the effect of interference 
on QoS.} The high demand for wide-area access 
to wireless data services combined with open access 
to lower frequency bands could create excessive congestion
in those bands leading to a ``tragedy of the commons".  
Indeed, this is one of the main arguments for
granting exclusive-use licenses for spectrum.

It is unclear {\em a priori} which of the preceding effects will dominate and how this depends on the amount of unlicensed spectrum and consumer demand.   
In this paper, we introduce a model to gain insight into
such questions.
More precisely, we consider a market in which
incumbent SPs compete for a common pool of consumers.
Each SP has an existing allocation of licensed spectrum,
and we evaluate the effect of introducing unlicensed spectrum 
as an additional resource. Any incumbent SP  as well as new entrants
may offer service in the unlicensed band in addition
to its licensed band,
modeling the fact that this band has a low barrier to entry. 
To capture congestion effects, we assume that consumers 
in a particular band experience a congestion cost that 
depends on the total number (or mass) of customers assigned 
to that band. All customers in a licensed band are served by
the associated SP, whereas the customers in the unlicensed
spectrum may be served by different SPs. Our goal is to determine how the additional unlicensed spectrum 
affects both social (total) and consumer welfare.   

Our analysis builds upon the framework for price competition 
in markets for congestible resources developed in the operations, 
economics and transportation literature; see, for example 
\cite{levhari1978duopoly,AH03,HTW05,AO07,AF07,Xiao07,JohariWR10} 
and the discussion at the end of this section.  
In this framework, customers request service from firms 
(SPs) based on a {\em delivered price} that depends on
the price paid for the service, announced by the firm,
and the congestion cost. The firms then set prices to maximize revenue. 
In prior work it is generally assumed that 
each firm only has access to resources for exclusive use.
The unlicensed spectrum in this paper can be viewed as an additional 
{\it non-exclusive} resource made available to each 
firm.\footnote{In this sense unlicensed spectrum is a 
{\it congestible public good}, e.g., see \cite{ Scotchmer85}.}

To model customer preferences, we assume that
the delivered price of a service is a linear combination 
of an SP's announced price and the congestion cost. 
We consider the following two cases: 
(1) a {\em homogeneous} customer population in which
all customers weight the congestion cost and announced price in the same way, 
i.e, all customers see the same delivered price;
and (2) a {\em heterogeneous} customer population in which
there are two user groups (``high-" and ``low-QoS") with 
different price-congestion trade-offs.
In the heterogeneous model, adding unlicensed spectrum
could conceivably cause the market to segment, namely,
by assigning users desiring higher (lower) QoS to
licensed (unlicensed) spectrum.  For that reason
adding the unlicensed spectrum might be expected to increase
efficiency. We observe, however, that congestion causes the social welfare 
and consumer surplus to exhibit relatively complicated behavior.

Our main results are summarized as follows:
\begin{itemize}
\item[1.)]  The social welfare depends on the amount of unlicensed spectrum 
that is added to the market. Adding an amount of unlicensed spectrum 
in a particular range, starting from zero,
can cause the social welfare to \emph{decrease}. 
\item[2.)] In the homogeneous model, consumer surplus is a non-decreasing
function of the amount of unlicensed spectrum.
\item[3.)] In the heterogeneous model, {\em both} SP profit 
{\em and} consumer surplus can decrease.
\item[4.)] In the heterogeneous model, the customer surplus 
can be a complicated, non-monotonic
function of the amount of unlicensed spectrum added. (There
can be many break points between which the customer surplus
increases, decreases, or stays the same.)
\end{itemize}

The first result is perhaps counter-intuitive, and is reminiscent
of \emph{Braess's paradox} (\cite{braess1968}) in transportation networks:
adding resources can decrease total system utility.
A key difference here is that this decrease is caused 
from price setting by the SPs rather than by the users as in~\cite{braess1968}. 
The explanation for this decrease is that the incumbent SPs, 
when faced with new competition from the unlicensed band, 
may have an incentive to raise (instead of lower) their prices,
depending on the amount of bandwidth. 
That facilitates a shift of traffic to the unlicensed band, 
where the associated interference externality is then shared 
with other SPs, and causes the overall welfare to decrease.
The second result implies that with homogeneous customers, 
any such loss in total welfare consists solely of the loss 
in the SPs' profits from serving fewer customers after raising their prices.

In the homogeneous model, prices change continuously 
as a function of the amount of unlicensed spectrum being added. 
In the heterogeneous model, an SP may have an incentive 
to increase its price discontinuously in order to switch 
from serving both high- and low-QoS customers to
serving a smaller number of high-QoS customers. 
This shifts more low-QoS customers to the unlicensed band 
increasing congestion there. Hence, when this switch happens, 
customer surplus {\em decreases} along with SP profits.
Furthermore, the surplus can be strictly smaller than
with no unlicensed spectrum. This is summarized by
the preceding third and fourth results.
Overall, these results suggest that adding new unlicensed spectrum 
to existing allocations can affect social and customer welfare
in complicated ways, and may have unexpected effects.

As alluded to previously, the general framework of competition 
with congestible resources has been studied in a number of different settings. 
In the context of service industries, \cite{levhari1978duopoly} 
and \cite{AH03} consider models of competition among firms whose customers 
experience a latency given by a queueing delay. 
As in our work, customers select firms based on a linear combination 
of latency and price, where the weights in this combination 
can be heterogeneous across the 
customers.\footnote{In \cite{levhari1978duopoly} each customer's 
weight is chosen from a distribution with continuous support, 
whereas in \cite{AH03} there are two customer classes as assumed here.}  
Other related work considers models in which firms commit to
a given latency and then incur a cost to meet this commitment based on the 
number of customers they attract. (See \cite{LeLi97} and \cite{AF07} 
for a survey of this area.) Here we do not allow for such commitments. 

Closer to our application is work motivated by 
communication and transportation networks.
For example, \cite{EngelFG99} considers models in which privately owned 
toll roads compete for customers (drivers) who select roads based 
on the delivered price, whereas \cite{AO07, AO07b, HTW05} 
are motivated by communication networks such as the Internet, 
where different links may be owned by different SPs, and again 
customers select links based on a delivered price. 
A theme in much of that work is to characterize the inefficiencies 
that occur due to oligopolistic competition in congested markets, 
compared to the outcome under a benevolent social planner. 
In contrast, here we focus on the impact of adding unlicensed spectrum 
on the outcome of such oligopolistic competition. 
There have been similar studies of efficiency loss 
in so-called {\it selfish routing} models
without pricing (e.g., \cite{roughgarden2002bad}), and
where prices are set by a benevolent manager 
(e.g., \cite{CDR03}). That class of models has also been extended 
to allow for investment on the part of SPs as well as pricing decisions
(e.g.,~\cite{Sun04,JohariWR10,AO09,Xiao07}.
Here we assume that any investment is a sunk cost and 
focus solely on the pricing behavior of SPs.

Most of the aforementioned work related to communication networks 
is motivated by wire-line networks as opposed to the wireless setting 
we consider. An exception is \cite{Sun04}, which considers price and 
capacity allocation between two wireless SPs, each with licensed spectrum. 
Users respond to the sum of a price and a congestion term 
that reflects the probability that a user's service is blocked. 
Another related model can be found in 
\cite{Maille01}, which studies price competition between 
licensed wireless SPs. 
Users respond to ``perceived prices," which depend on congestion 
as well as an announced price, but the relationship is not linear
as in our model. Other work on competition among wireless SPs 
focuses on different issues such as the impact of auction design 
on competition \cite{cramton2011using}, the effect of 
roaming agreements and termination 
charges \cite{LaffontTiroleBook,ArmstrongWright2009}, 
and the impact of customer switching costs \cite{shi2006price}.

The rest of the paper is organized as follows: 
Section~\ref{sec:model} introduces our model, and  
Section~\ref{sec:result} studies and compares social welfare 
and consumer surplus within this framework.
Conclusions are given in Section~\ref{sec:conclude}, and 
some proofs and numerical calculations are provided in the appendix.


\section{The Model}\label{sec:model}
We present our heterogeneous model assuming two classes of consumers, 
each class having a different sensitivity to delay. The homogeneous model 
is then presented as a special case.

\noindent
{\bf Service Providers}\\
We assume a set of incumbent SPs, each of which has its own licensed band,
and a set of new entrants, which do not have licensed spectrum
and must use the unlicensed band to offer service.
(For tractability our analysis will primarily assume a single incumbent.)
Each SP competes for customers by announcing a price
for using its licensed band (if it is an incumbent) and another price for using 
the unlicensed band. The SP then serves all customers who accept their posted  price. Suppose an SP $i$ sets price $p_i$ for service in its licensed band, price $p^w_i$ in the unlicensed band,
and serves $x_i$ and $x^{w}_i$ customers in those bands, respectively.   
Then $i$'s revenue is given by $\pi_i=p_i x_i+p_i^w x_i^w$. 
(Here $w$ stands for ``white space''.)

There is a congestion externality due to the interference
suffered by customers in both the licensed and unlicensed bands. 
If SP $i$'s licensed band serves a mass of customers $x_i$,
then each customer served in this band experiences a 
{\it congestion cost} $l_i(x_i)$, which depends on the
bandwidth and the technology deployed by SP $i$.
The congestion cost in the unlicensed band, however, depends on
the {\em total} mass of customers served in that
band by {\em all} SPs. Specifically, letting $x_i^w$ be the mass of
customers served by SP $i$ in the unlicensed band, the congestion
cost for customers served in the unlicensed band is
$g(X^w)$, where $X^w=\sum_{i \in \mathcal{N}} x_i^w$.\footnote{ \highlight An implicit assumption here is that all customers require the same amount of service (on average) from the network, so that congestion only depends on the number of customers in a band and not their individual demands.} The congestion cost $g(X^w)$ also depends on the bandwidth of the
unlicensed band and the technology, which we assume is the same for all SPs. 
In this paper we consider the case where $l_i$ is fixed
and $g$ varies according to the available bandwidth of the unlicensed spectrum (in a manner to be specified).

Throughout the paper we assume that for a given bandwidth, all congestion costs are monotonically increasing and convex
functions of the load (mass of customers served). {\highlight A simple example of such a congestion cost, which we use later, is $\frac{x}{C}$ where $C$ is the bandwidth. This makes the congestion cost proportional to the time for each customer to send a fixed length file, assuming the bandwidth is equally divided among the active customers, the number of which is proportional to $x$. Another example, as in \cite{AH03}, is to assume the congestion cost is proportional to the average delay in a queue where the arrival rate is proportional to the mass of customers in the band.}

{\highlight  In our  model, an incumbent SP offers two distinct
services, one in the licensed band and one in the unlicensed band, 
with a separate price and congestion cost for each.
While reasonable, there could be other pricing policies offered by an incumbent. For example, 
it could offer a single service (with a single price) and then split its customers between the 
two bands so that they experience the average latency of the two bands over time. 
We do not consider such variations in part to keep the analysis from 
becoming overly complicated. Also, if there are multiple entrants using 
the unlicensed band, then it can be seen from our analysis that new entrants 
may congest the unlicensed band so that incumbents would not want to 
utilize that band in this way.}

\noindent
{\bf Customers}\\
{\highlight 
As in much of the prior literature on competition with congestible resources, 
we consider a unit mass of infinitesimal customers who 
choose an SP based on the \textit{delivered price}, 
which is the weighted sum of the price
announced by an SP and the congestion cost she experiences 
when served by that SP. We consider a simple model for heterogeneous customers
in which there are two different classes of customers,
delay sensitive (high Quality of Service) and 
delay insensitive (low Quality of Service). The classes are distinguished by how much 
weight they place on the congestion cost, when determining their delivered price.}\footnote{\highlight The assumption that customers care only about the weighted sum of congestion and price, where the weight differentiates the classes, is made for tractability. We also only consider two classes of customers for simplicity. We conjecture that in a more general model with more customer classes (or where the customer utility is a nonlinear function of congestion and price, which may vary across different classes of customers) our qualitative results continue to hold. This is because with  just two categories of service, an SP can partition customers into at most  two classes and offer at most two different prices.}

Specifically, for a customer of type $t\in \{h,l\}$ (high, low)
served by SP $i$, the \textit{delivered price} in the licensed band 
is $p_i+\lambda_tl_i(x_i)$, where $\lambda_t$ is the relative weight,
and $\lambda_h > \lambda_l$. The delivered price in the unlicensed band 
is $p_i^w+\lambda_t g(X^w)$.
Customers within each class (high/low) choose the SP and 
type of service (licensed or unlicensed) with the lowest delivered price. 
When facing the same delivered price from
multiple SPs, customers randomly choose one of the SPs. 

The demand for services from the two classes is given by
two downward sloping demand functions $D_h(p)$ and $D_l(p)$ 
with inverse functions $P_h(q)$ and $P_l(q)$, 
respectively.\footnote{In other words, $P_t(q)$ gives the
maximum delivered price for which $q$ customers of type $t$ 
would be willing to purchase service.} As is standard,  we assume that $P_h$ and $P_l$ are concave in $q$.
A special case of this {\em heterogeneous} customer model
is the {\em homogeneous} model in which $\lambda_h = \lambda_l$,
so there is only one type of customer. 
(Alternatively, one of the demand functions can be set to zero.) In this case, we denote the overall inverse
demand as simply $P(q)$.

\noindent
{\bf Pricing Game and Equilibrium}\\
We consider a game in which SPs first simultaneously announce
prices on licensed and unlicensed bands. Customers then choose 
SPs based on the delivered price. In this section we characterize
the corresponding equilibrium along with the associated
social welfare and consumer surplus.

Let $x_i^h, x_i^l$ be the number (measure) of customers of each type
that receive service from SP $i$ in the licensed band.
Similarly, let $x_i^{wh}, x_i^{wl}$ be the number of customers 
of each type served by SP $i$ in the unlicensed band. 
Thus, $x_i=x_i^h+x_i^l$ and $x_i^w=x_i^{wh}+x_i^{wl}$. 
We assume each customer is infinitesimally small 
and adopt the notion of Wardrop equilibrium to characterize 
how demand is allocated (\cite{Wardrop52}). 
Namely, given a price vector $(\mathbf{p},\mathbf{p}^w)$, 
the non-negative demand vector 
$(\mathbf{x^h},\mathbf{x^l},\mathbf{x^{wh}},\mathbf{x^{wl}})$ 
induced by $(\mathbf{p},\mathbf{p}^w)$ must satisfy in the licensed bands:
\begin{equation}\label{eq:wardrop_lic}
\begin{split}
p_i+\lambda_tl_i(x_i) &=P_t(Q_t),\qquad \text{if}\,\,\,\,x^t_i>0, \; t\in\{h,l\}\\
p_i+\lambda_tl_i(x_i) &\geq P_t(Q_t),\qquad \text{if}\,\,\,\,x^t_i=0, \; t\in\{h,l\}
\end{split}
\end{equation}
and in the unlicensed bands:
\begin{equation}\label{eq:wardrop_unlic}
\begin{split}
p_i^w+\lambda_tg(X^w)&=P_t(Q_t),\qquad \text{if}\,\,\,\,x^t_i>0, \; t\in\{h,l\} \\
p_i^w+\lambda_tg(X^w)&\geq P_t(Q_t),\qquad \text{if}\,\,\,\,x^t_i=0, \; t\in\{h,l\}.
\end{split}
\end{equation}
Here, $Q_t=\sum_i{(x^t_i+x^{wt}_i)}$ for $t\in \{h,l\}$ is the total number of customers of type $t$ served in the market. 

\begin{remark}
It is straightforward to show that given a price vector, 
the corresponding demand vector satisfying the above conditions 
always exists and is the solution to a convex 
program. (See, for example, \cite{AO07} for a proof 
of a similar result without the addition of an analogous unlicensed resource.)
\end{remark}
We next define the equilibrium notion we will use for the overall game.

\vspace{3mm}
\begin{definition}\it
A pair $(\mathbf{p},\mathbf{p^w})$ and
$(\mathbf{x^h},\mathbf{x^l},\mathbf{x^{wh}},\mathbf{x^{wl}})$ is a pure strategy Nash
equilibrium if $(\mathbf{x^h},\mathbf{x^l},\mathbf{x^{wh}},\mathbf{x^{wl}})$ satisfies
equation (\ref{eq:wardrop_lic}) and (\ref{eq:wardrop_unlic}) given $(\mathbf{p},\mathbf{p^w})$,
and no SP can increase its revenue by changing its prices.
\end{definition}
\vspace{3mm}

\noindent
{\bf Social Welfare and Customer Surplus}\\
Next, we define the notions of \emph{social welfare}  and  \emph{customer surplus} in this setting.

\begin{definition}\label{defn:sw}
\it Let $(\mathbf{x^h},\mathbf{x^l},\mathbf{x^{wh}},\mathbf{x^{wl}})$
be the demand vector
induced by some price vector $(\mathbf{p},\mathbf{p^w})$ 
according to (\ref{eq:wardrop_lic}) and  (\ref{eq:wardrop_unlic}). 
Then the \emph{social welfare} is given by
\begin{equation}\label{eqn:defn_ss}
\begin{split}
SW= &\int_{0}^{Q_h}P_h(q)dq+\int_{0}^{Q_l}P_l(q)dq\\
  &-\sum_{i\in\mathcal{N}}\lambda_hx_i^hl_i(x_i)
-\sum_{i\in\mathcal{N}}\lambda_lx_i^ll_i(x_i)-\lambda_hg(X^w)X^{wh}-\lambda_lg(X^w)X^{wl},
\end{split}
\end{equation}
where $X^{w}=X^{wh}+X^{wl}$, and $X^{wh}$ and $X^{wl}$ 
are the number of high and low customers 
in the unlicensed band, respectively, and
$Q_h=\sum_{i\in\mathcal{N}}x_i^h+X^{wh}$ and $Q_l=\sum_{i\in\mathcal{N}}x_i^l+X^{wl}$ are the total number of high and low customers served in the market.
\end{definition}

\begin{definition}\label{defn:cmsur}
\it Given price and demand vectors, let $\Delta_h$ and $\Delta_l$ be the 
resulting delivered price for high and low customers, respectively. 
Then the \emph{customer surplus} is given by
\begin{equation}\label{eqn:defn_cs}
CS=\int_{0}^{Q_h}(P_h(q)-\Delta_h)dq+\int_{0}^{Q_l}(P_l(q)-\Delta_l)dq,
\end{equation}
where $Q_h$ and $Q_l$ are defined in Definition \ref{defn:sw}.
\end{definition}

\begin{figure}[htbp]
\centering
\includegraphics[ width=5in]{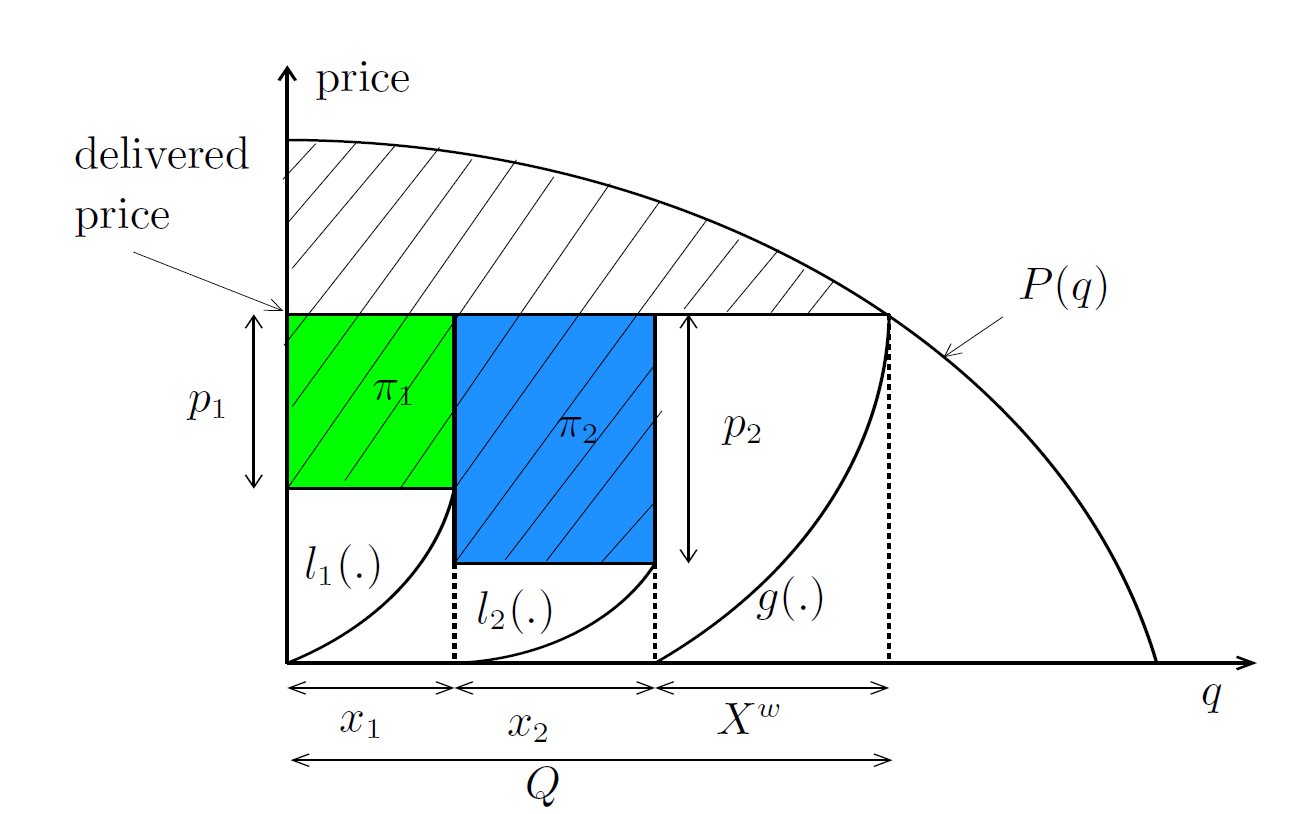}
\caption{\highlight  Illustration of pricing game with two SPs, 
homogeneous customers and unlicensed spectrum. The quantities
$x_1, x_2$ and $X^w$ are the amount of customers served by 
SP1, SP2 and unlicensed spectrum, respectively.
The areas of the rectangles $\pi_1, \pi_2$ are the revenues of SP1 and SP2, respectively.
The area of the region between the delivered price and the inverse demand curve $P(q)$ 
is the consumer surplus, and the dashed area is total social welfare.} 
\label{fig:example}
\end{figure}

Figure~\ref{fig:example} shows an example of the pricing game, 
where there are two incumbents with latency functions 
$l_1(\cdot)$ and $l_2(\cdot)$ in their respective licensed bands. 
The latency function of the white space is given by $g(\cdot)$. 
The customer population is homogeneous with the 
single inverse demand curve $P(q)$.
At an equilibrium in which all bands are used, 
the Wardrop conditions imply that the delivered prices 
across the licensed and unlicensed bands are the same. 
That is,
$$
l_1(x_1)+p_1=l_2(x_2)+p_2=g(X^w)+p^W.
$$
Figure~\ref{fig:example} shows that the price charged in
the unlicensed band, $p^W=0$.
We will show in the next section that this is true at any Nash equilibrium.
The revenue of SP $i$ is $\pi_i=p_i x_i$, $i=1,2$,
and corresponds to the area of the indicated rectangle.
The social welfare is the shaded area shown in the figure. 
The consumer surplus is the shaded area above the horizontal line 
at the equilibrium delivered price, and is equal to the 
social welfare minus the revenue of all SPs.


\section{Main results}\label{sec:result}

We first show that  the announced price in equilibrium in the unlicensed band is the marginal cost,  which we assume to be zero. We then use this result to investigate  the effect of adding unlicensed spectrum on the social welfare  and consumer surplus.

\subsection{Equilibrium Price in Unlicensed Spectrum}
Let $\mathbf{p}^* $ and $(\mathbf{x^h}^{*},\mathbf{x^l}^{*})$ denote the 
equilibrium price and demand vectors in the licensed bands, respectively,
and $\mathbf{p}^{w*} $ and $(\mathbf{x^{wh}}^{*},\mathbf{x^{wl}}^{*})$ 
denote the corresponding equilibrium prices and demands in 
the unlicensed band. 
Let $Q_t^*$ be the total number of type $t$ customers served in the equilibrium.
Also recall that $g(\cdot)$ is the congestion cost in the unlicensed band;
$g(0)>0$ then represents some fixed cost experienced by all customers
in that band,  e.g., due to sharing with other systems and services.

\begin{lemma}\label{prop:ws_zeroprice}
Given at least two SPs in a market with unlicensed spectrum, 
at any equilibrium either $\mathbf{p}^{w*} = \mathbf{0}$,
with at least two SPs serving a positive mass of customers, 
or no customers are served in that band. 
Furthermore, if no customers are served, then $g(0) \geq P_t(Q_t^*)$ 
for both types of customers. 
\end{lemma}

\begin{remark}
This result can be easily extended to a scenario where each SP has a non-zero  marginal cost for providing the unlicensed service. 
In that case, the equilibrium unlicensed price will be the marginal cost.  
The qualitative result is the same, and thus for simplicity, we assume 
the marginal costs are zero. (This can be interpreted as having zero-cost devices.)
\end{remark}

This result follows from the assumption that all customers
using the unlicensed spectrum experience the same congestion cost.
Hence the SPs with the lowest price capture the entire market for a given
customer class. If there are at least two SPs only serving 
customers in the unlicensed band, competition will then drive the prices to zero.   However, if SPs competing in the unlicensed band also offer 
services in licensed spectrum, then lowering the price for 
unlicensed service can lower their revenue in the licensed band.  
We next show that even in this case prices will be driven to zero.
If no one offers service in the unlicensed band, then
the delivered price in the unlicensed band ($g(0)$) must exceed that 
in the licensed bands ($P_t(Q_t^*)$). If the congestion cost
in the unlicensed band satisfies $g(0) = 0$, then
that scenario cannot occur.

\vspace{5mm}
\begin{proof}{\bf Proof of Lemma~\ref{prop:ws_zeroprice}:}
Assume that in equilibrium $\mathbf{p}^{w*} \neq \mathbf{0}$ and some SP
serves a strictly positive mass of customers. Call an SP \emph{active} 
if in equilibrium she sets a positive price that results in a
strictly positive quantity of customers. The Wardrop equilibrium conditions imply that all active SPs in the unlicensed band must charge the same price, $p^{w*} > 0$.

Furthermore, in equilibrium if one SP is active in the unlicensed band, 
then all SPs must be active in that band. Otherwise, an inactive SP 
could increase its revenue by charging the same price in the unlicensed band as the active SP. That would increase the number of customers it serves without decreasing revenue from the licensed band. Thus, it follows that in any equilibrium either all SPs are active  and charge the same price $p^{w*} >0$, or no SP serves a  positive mass of customers in the unlicensed band. 
The latter case contradicts our assumption, hence we focus on the first case.

Given that all SPs are active and charge the same price in the unlicensed band, consider the effect of an SP $i$ dropping her price 
by $\epsilon$. It must be that before dropping her price 
she is serving $x_i^w <X^w$. Since all customers experience 
the same latency, all $X^w$ customers will switch to SP $i$.
Furthermore, some customers currently on licensed bands 
will also switch to $i$'s unlicensed service.  

It follows that if SP $i$ is an entrant, then she can significantly 
increase her revenue by dropping the price by a small amount.   
However, if SP $i$ is an incumbent, dropping her price in the 
unlicensed band by $\epsilon$ can decrease her revenue 
in the licensed spectrum.  Nevertheless, if $\epsilon$ 
is small enough, we next show that SP $i$ still increases its revenue. 
Let the equilibrium price in the licensed band of SP $i$ be $p^*_i$, 
and assume it keeps this price. Suppose that after dropping the price 
in the unlicensed band by $\epsilon$ the customer mass of SP $i$ 
in the licensed band is reduced by $\delta_{x_i}$.  
The overall change in $i$'s profit $\pi_i$  is given by
\begin{align*}
\delta_{\pi_i}& \geq (p^{w*}-\epsilon)X^{w*}-(p^{w*}x_i^{w*}+\delta_{x_i}p^*_i)\\
&= p^{w*}(X^{w*}-x_i^{w*})-\epsilon X^{w*}-\delta_{x_i}p^*_i.
\end{align*}
Since $\lim_{\epsilon\rightarrow 0}\delta_{x_i}=0$, 
there exists a sufficiently small
$\epsilon>0$ such that $\delta_{\pi_i}$ is strictly positive. 
Thus, decreasing the price in the unlicensed band 
is a profitable deviation for SP $i$. This contradicts the initial assumption.
Therefore, $p^{w*}_i=0$ for every SP $i$ serving customers in the unlicensed 
band. Furthermore, for this to be an equilibrium at least two SPs 
must serve customers at this price.\footnote{SPs would be indifferent 
between announcing a price of zero or an arbitrarily high price 
for unlicensed service, since in either case their revenue is zero.} 

Finally, to prove the last part of the lemma, suppose that no SP is 
serving customers in the unlicensed band in equilibrium, 
but $g(0) < P_t(Q_t^*)$.  An incumbent SP could then increase its revenue
by offering unlicensed service. Hence, any incumbent SP that deviated to offer unlicensed service would be the sole provider of unlicensed service. 
Suppose that such a provider maximizes its revenue 
across the licensed and unlicensed bands while keeping 
the total number of customers served fixed 
(so as to not change the delivered price). 
Inspecting the optimality conditions of this problem, 
it can be seen that the incumbent can always increase
its revenue by using both bands, leading to a contradiction.
\qed

\end{proof}

\subsection{Social Welfare}
We now analyze how the social welfare varies with the amount of unlicensed spectrum added. 
We start with homogeneous consumers and show that for a class of demand and latency functions, 
social welfare first {\em decreases} and then increases as more unlicensed spectrum is added. 
We then show that with heterogeneous customers the total welfare 
can vary in a more complicated manner.

\subsubsection{Homogeneous model:}\label{sec:homogen}
Suppose a single incumbent (monopoly) initially operates in a licensed band, and an unlicensed band becomes available. Both new entrants and the incumbent can then offer service in the unlicensed band. (The new entrants 
only use the unlicensed band.)  In this section we focus on the homogeneous model and so without  loss of generality we set $\lambda_l=\lambda_h=1$ and drop the  customer type from the notation. 

As in \cite{AO07}, we consider the case where all users 
have the same valuation $W$ for service
(i.e., users are homogeneous not only in how they weight 
congestion and price, but also in how they value the service).  
This corresponds to the inverse demand $P(x)$
having constant value $W$ for $0\leq x\leq 1$ and then 
dropping to zero for $x>1$.  
Customers choose an SP based on delivered price as long 
as it is at most $W$. We focus on the case where $W$ is such 
that prior to adding the unlicensed band not all customers 
are served by the incumbent. That can occur because
the incumbent is a monopolist and so has an incentive to 
limit supply to extract a higher price, and also because
serving all customers would produce an excessively high congestion cost.

We will assume that the congestion cost (latency) in each band is linear.
That is, the latency in the licensed band is
$$ l(x)=T_1+bx,$$
where $b>0$ and  $0\le T_1\le W$, and
the latency in the unlicensed band is $$g(x)= T_2 + \alpha_C x,$$
where $0\le T_2\le W$ and $C\ge 0$ denotes the unlicensed bandwidth.
We assume that $\alpha_C>0$ is decreasing in $C$; 
when $C=0$ (no unlicensed spectrum), $\alpha_0 =\infty$, 
and as $C\rightarrow \infty$, $\alpha_C \rightarrow 0$.
The offsets $T_1, T_2$ can be interpreted as
fixed costs for connecting to the SP. We also assume
$$
g(1)>l(0) \text{ and  } l(1)>g(0),
$$
that is, the congestion cost of  serving the whole market in one band exceeds the fixed cost of connecting in the other.

The following theorem describes what happens when the incumbent's bandwidth, 
corresponding to the coefficient $b$ of the congestion cost, is fixed 
and we increase $C$, the bandwidth allocated to the unlicensed band.

\begin{figure}[htbp]
\centering
\psfrag{Social}{$S(C)$}
\psfrag{S}{$S$}
\psfrag{C1}{$C_1$}
\psfrag{C2}{$C_2$}
\psfrag{UnCap}{$C$}
\includegraphics[height=1.8in]{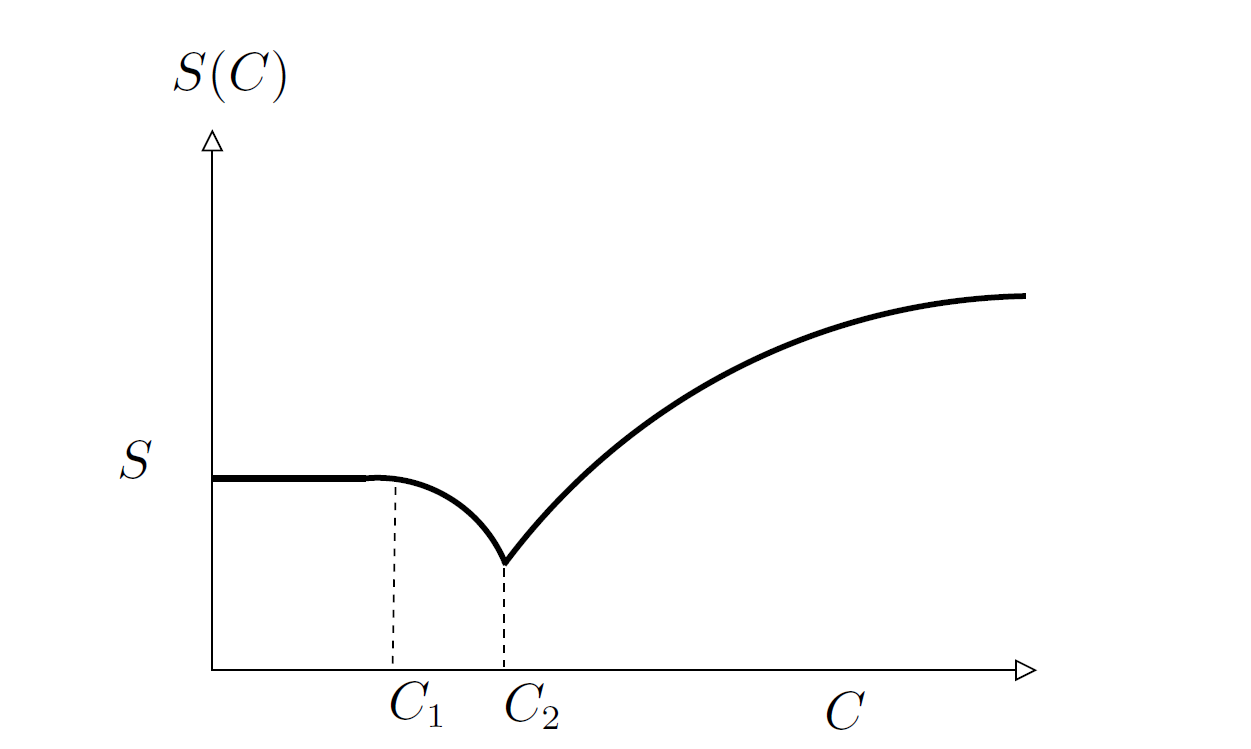}
\caption{Social welfare as a function of the bandwidth of unlicensed spectrum for a single incumbent and homogeneous users.}
\label{fig:transition}
\end{figure}

\begin{theorem}\label{thm:ss_wws}
Consider an incumbent SP with licensed spectrum that does not 
serve all of the demand. If an amount of unlicensed spectrum $C$ is added then:
\begin{itemize}
\item[\bf (i)] For every  $C\ge 0$ there is a unique equilibrium.
\item[\bf (ii)] 
The social welfare at an equilibrium, $S(C)$, exhibits the following behavior.
There exist $0<C_1<C_2\le \infty$ such that $S(0)=S(C_1)>S(C_2)$ and
\begin{itemize}
\item $S(C)= S(0)$ for $0\le C\le C_1$; 
\item $S(C)$ is monotone decreasing for $C_1\le C\le C_2$; 
\item $S(C)$ is monotone increasing for $C \ge C_2$.
\end{itemize}
\end{itemize}
\end{theorem}

Figure~\ref{fig:transition} illustrates the behavior of $S(C)$ as specified in this theorem.
The formal proof is given in Appendix~\ref{app:theo1}. 
What follows is an informal explanation of this result.

\begin{figure}[htbp]
\centering
\psfrag{p}{$p_1^*$} \psfrag{2p}{$2p_1^*$}
\psfrag{p'}{$p_1$}
\psfrag{c}{$1-x_1^*$}
\psfrag{1+c-x}{$X^w=1-x_1$}
\psfrag{lx}{$l(x)$}
\psfrag{x}{$x_1$}
\psfrag{gx}{$g(x)$}
\psfrag{Dx}{$D(x)$}
\psfrag{1}{$x_1^*$}
\psfrag{y}{$X^w$}
\psfrag{W}{$W$}
\psfrag{W'}{$P$}

\psfrag{de}{delivered price}
\psfrag{cus}{customers}

\includegraphics[height=3.2in]{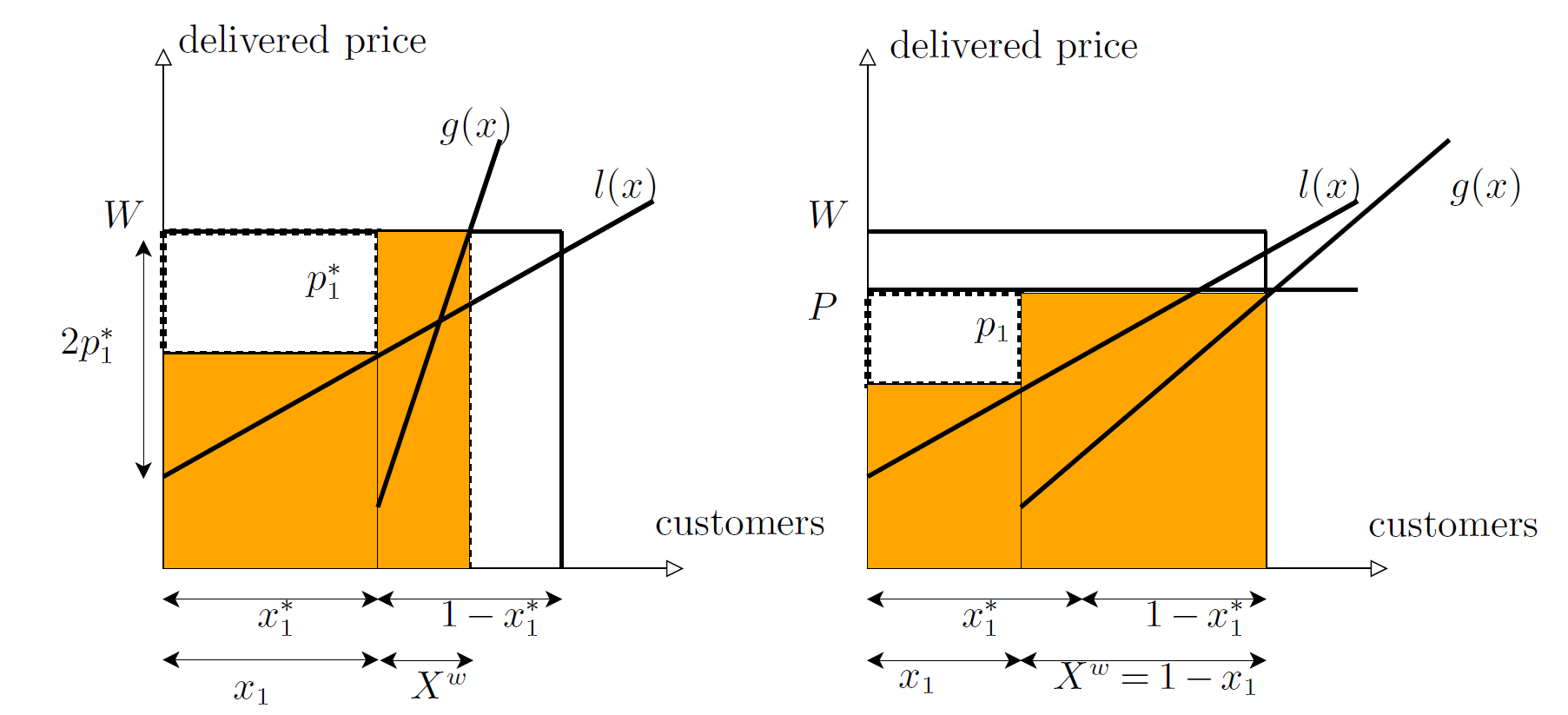}
\caption{Illustration of adding a relatively small amount of  unlicensed spectrum (left) and a relatively larger amount (right).}
\label{fig:addingwhitespace}
\end{figure}

{\highlight 
Without unlicensed spectrum, it can be seen that the incumbent extracts 
all of the welfare with the assumed inverse demand, i.e., 
the delivered price is $W$. Furthermore, by assumption, when $C = 0$, 
the incumbent's price $p_1^*$ is such that the number of customers 
it serves $x_1^* < 1$.  Figure~\ref{fig:addingwhitespace} illustrates 
the effect of adding a relatively small amount of unlicensed spectrum (left figure) and a larger value (right figure) to such a market.  In the left figure, adding the unlicensed band enables $X^w$ additional customers to be served, without changing the fraction of customers served in the licensed band ($x_1^*$).
The congestion cost in the unlicensed band $g(X^w) = W$,
which is also the delivered price due to Lemma \ref{prop:ws_zeroprice},
so that the consumer surplus is zero and the total welfare remains equal to the 
incumbent's revenue, i.e., 
$S(C) = p_1^* x_1^*$.
Hence, adding the unlicensed spectrum in this scenario affects neither
the prices (announced and delivered) nor the social welfare. This is due to the high congestion cost in the unlicensed band.

As $C$ increases, the slope of the latency function $g(x)$ decreases
until the latency curve $g(x)$ passes through the corner point of the inverse demand
so that $g(1-x_1^*) = W$ and all customers are served. This corresponds to $C=C_1$.
Increasing $C$ further decreases the delivered price, as shown in the right figure.
Better service in the unlicensed band then attracts customers away from
the licensed band so that the consumer welfare increases and the SP revenue decreases.
The incumbent need not respond to this erosion in share
with a price cut. In fact, the incumbent might benefit from a price increase. 
That would drive even more customers into the unlicensed band, worsening 
the service quality there. Customers that remain in the licensed band pay more, 
but they receive  a higher quality of service and have no incentive to use the unlicensed band.
The number of customers consuming lower quality service then increases, 
which increases the overall congestion cost and reduces social welfare.
As $C$ continues to increase, the delivered price continues to fall
and consumer welfare rises until it dominates the loss in SP revenue
so that total welfare starts to increase (at $C=C_2$).

}

\begin{remark}
{\highlight 
It is shown in Appendix~\ref{app:example} that 
for the class of demand and latency functions assumed here,
adding the unlicensed spectrum can reduce the social welfare 
by as much as $62\%$.
The same qualitative result is also observed  with multiple competing incumbents and a more general class  of demand functions; see Appendix~\ref{multiple_incumbents} for a numerical example. 

In Appendix~\ref{app:general} we show that the loss in social welfare as
stated in Theorem~~\ref{theo:general} also applies to a broader class of
latency functions. Specifically, we show that if the 
latency functions satisfy certain convexity conditions so that
the SP revenue is concave in the number of customers it serves, then
there is a range of parameter such that adding unlicensed spectrum 
decreases social welfare.  An example is when
the congestion cost of the unlicensed band is linear and  
the congestion cost of the licensed band is increasing convex.  
A similar argument applies to this scenario. Namely, when faced with competition from a small amount of unlicensed spectrum
the incumbent will increase its announced price, which
induces some customers to switch from the licensed  
to unlicensed band increasing congestion and reducing total welfare.

}

\end{remark}


\subsubsection{Heterogeneous Model:}
{\highlight With two customer types, one might expect that adding unlicensed spectrum increases efficiency since the market may be able to sort the high/low customers into the proprietary/unlicensed bands. We instead show that with heterogeneous customers, the social welfare again does not generally increase monotonically with  increasing unlicensed bandwidth, and exhibits more complicated behavior when compared with the homogeneous model.}

We again consider the scenario in which an unlicensed band is added
to a licensed band used by a single incumbent (monopolist).
The incumbent, as well as new entrants, can then offer services in the unlicensed band.  As in Section~\ref{sec:homogen}, all customers of the same type $t$  have a common valuation $W_t$ for service, but now we allow two different valuations
for the two classes. That is, the inverse demand $P_t(q)$ for each type $t$ 
is a constant $W_t$ for $0\leq q\leq Q_t$ and then drops to zero 
for $q\geq Q_t$, where $Q_t$ is the total mass of type $t$ customers.
The SP then serves customers provided that its delivered price is at most $W_t$. We also assume $W_h>W_l$ and $Q_h<Q_l$,
i.e., the high-type customers have higher valuations for the
service and there are more low- than high-type customers.

\begin{figure}[htbp]
\centering
\psfrag{Social}{$S(C)$}
\psfrag{C1}{$C_1$}
\psfrag{C2}{$C_2'$}
\psfrag{C2p}{$C_2$}
\psfrag{C3}{$C_3$}
\psfrag{C4}{$C_4$}
\psfrag{UnCap}{$C$}
\includegraphics[height=2.5in]{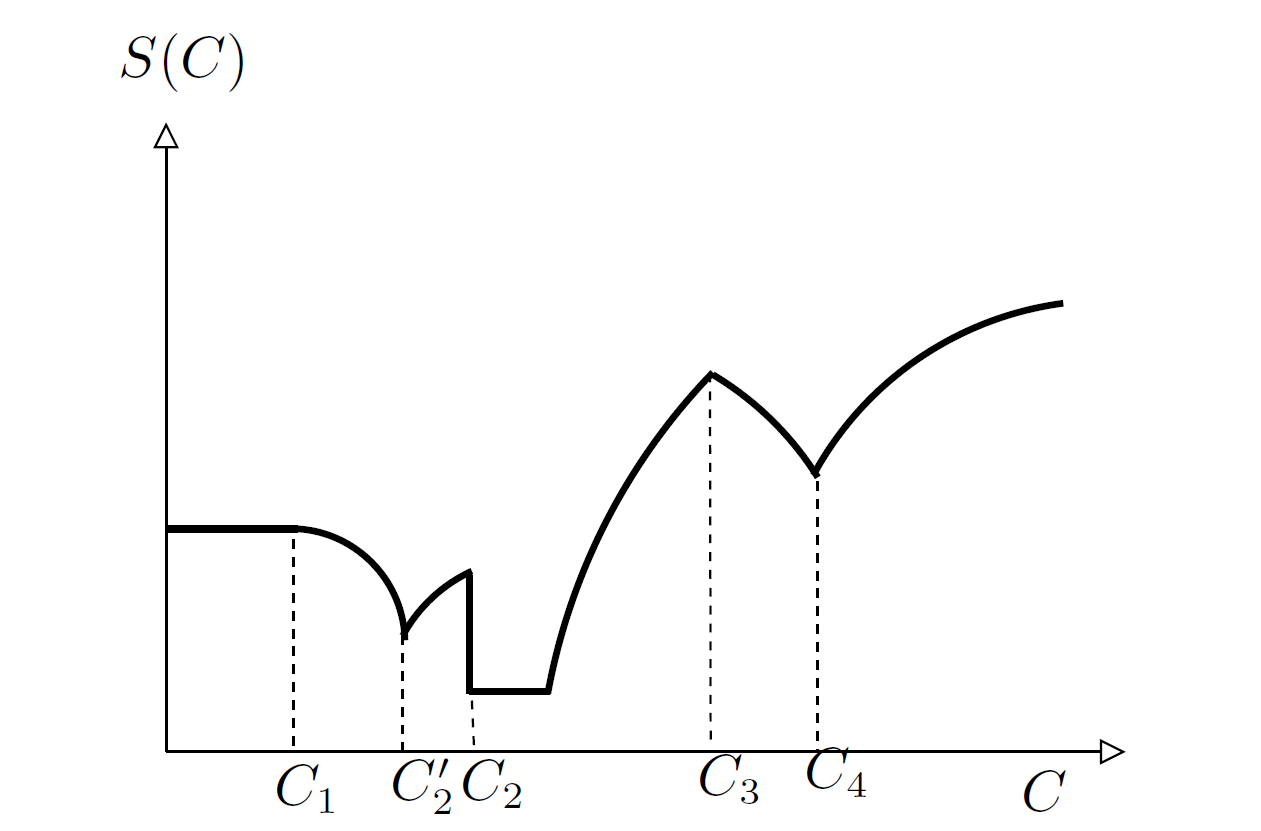}
\caption{Example of social welfare as a function of the unlicensed bandwidth for the heterogeneous model.}
\label{fig:transition_heter}
\end{figure}

Figure~\ref{fig:transition_heter} shows an example of social welfare versus
unlicensed bandwidth $C$ for the heterogeneous model. The example is described in Appendix~\ref{app:heterogeneous}. 
Namely, when $0\le C\le C_2$ the incumbent serves both types of consumers
and the shape of the curve is similar to that for the homogeneous model. 
When $C=C_2$ there is a sudden drop in the social welfare. 
At this threshold, the incumbent increases the price {\em discontinuously}
in an attempt to serve only high-type customers.
This causes all low-type customers to switch to the unlicensed band. 
When $C$ increases from $C_2$ to $C_3$, congestion in the unlicensed spectrum decreases, 
which increases the total social welfare. When $C>C_3$, the unlicensed band
starts to compete with the licensed band again and we get a similar effect 
as in the homogeneous model: social welfare decreases until $C=C_4$, 
and then starts to increase again.

The sudden increase in the price at $C=C_2$ and drop in the social welfare 
will be analyzed in the next section. Since the unlicensed band competes with the licensed band,  adding unlicensed spectrum can only decrease the revenue of the monopolist. We will see that the drop in social welfare at $C=C_2$ is caused by a decrease in {\em consumer surplus} as well as the monopolist's revenue.


\subsection{Consumer Surplus}\label{sec:consumer}
In this section we analyze the change in customer surplus when unlicensed spectrum 
is added to an existing market for wireless services offered in licensed spectrum. 
We show that when customers are homogeneous, adding unlicensed spectrum can never 
increase the delivered price, so that consumer surplus cannot decrease.  
However, an important difference emerges with the heterogeneous model: adding
unlicensed spectrum induces the SP to serve only high-type customers.
Specifically, suppose that the incumbent SP serves both high- and low-type 
customers with no unlicensed band.
If a sufficiently small amount of unlicensed spectrum is added,
then the incumbent will continue to serve both classes of customers. 
However, as the unlicensed bandwidth increases,
the incumbent will raise the price to serve only high-type customers causing  
a drop in customer surplus. We show that this effect is \emph{always} present in heterogeneous models  for general (concave, non-increasing) demand curves and (convex, increasing) latency functions.

Our results are stated in the next two theorems.
The first pertains to the homogeneous model.

\begin{theorem}\label{theo:homogeneous}
Consider a single SP with licensed spectrum and a set of 
homogeneous customers, i.e. $\lambda_l=\lambda_h$. 
Let $\Delta_0$ and $\Delta_1$ be the delivered equilibrium prices before and after 
the unlicensed spectrum is introduced, respectively. Then $\Delta_0\geq\Delta_1$.
\end{theorem}

\begin{figure}[htbp]
\centering
\psfrag{a}{$a_1$} \psfrag{a1}{$a_1$}
\psfrag{x}{$x_3$} \psfrag{x1}{$x_1$}
\psfrag{l(x)}{$l(x)$}
\psfrag{lw}{$l_w(x)$}
\psfrag{D}{$\Delta_0$} \psfrag{D1}{$\Delta_1$}
\psfrag{A}{$A$}
\psfrag{B}{$B$}
\psfrag{A'}{$A'$}
\psfrag{B'}{$B'$}
\psfrag{p}{$p$} \psfrag{p1}{$p_1$}
\includegraphics[height=2.3in]{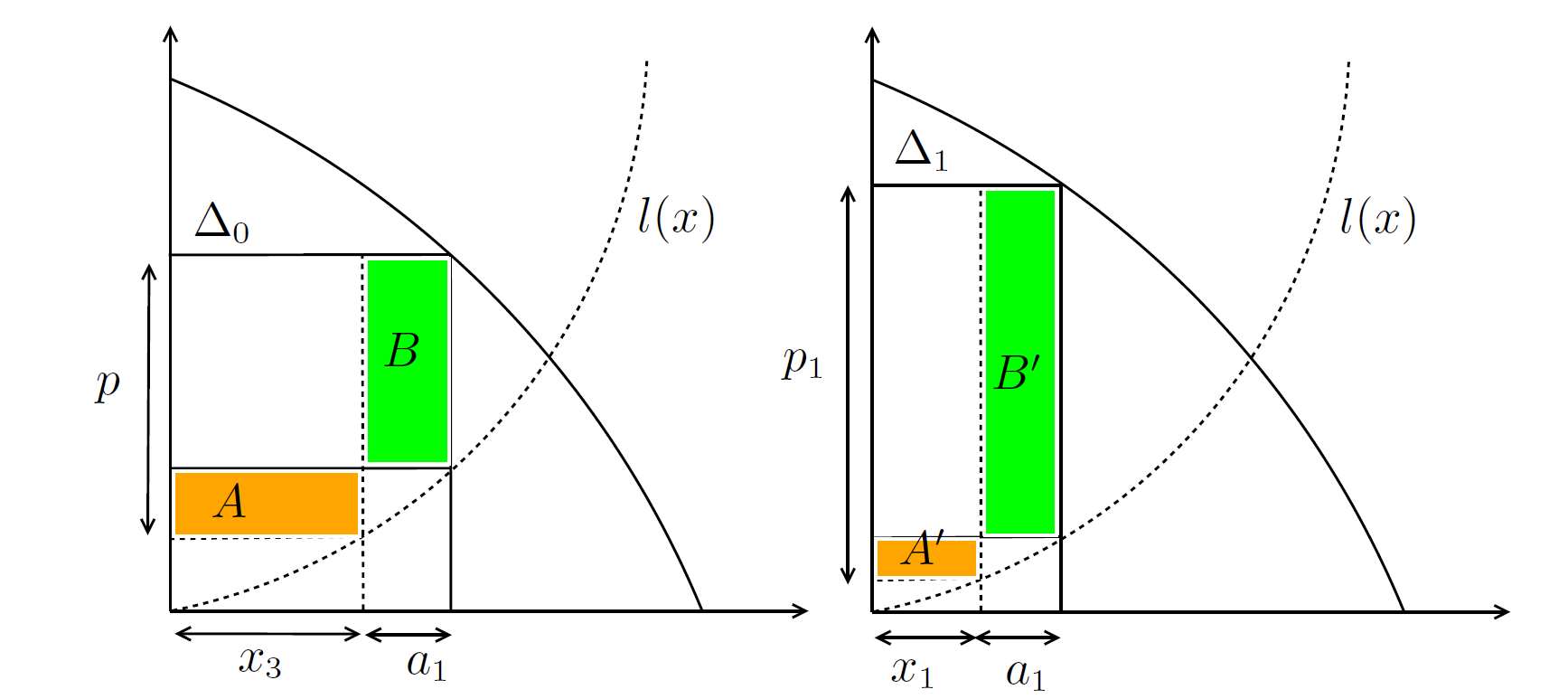}
\caption{Illustration of the revenue differences 
in the proof of Theorem~\ref{theo:homogeneous}.}
\label{fig:123}
\end{figure}

\noindent
{\bf Proof:}
The proof is by contradiction. We assume homogeneous customers, a single incumbent,
and general demand and congestion functions.
Let $\Delta$ be the delivered price, and let $a$ be the 
corresponding number of customers in the unlicensed band and 
$x$ and $p$ be the corresponding number of customers 
and price of the licensed band, respectively.
(See Figure~\ref{fig:123}.)

We will consider the following three cases: 
({\it i}) There is no unlicensed spectrum and the incumbent maximizes its revenue. 
For this case, let $\Delta_0$ be the delivered price, so that $a_0=0$ and $\Delta_0=l(x_0)+p_0$. 
({\it ii}) A fixed amount of unlicensed spectrum is introduced 
and the provider charges a price to maximize revenue under competition 
with the unlicensed spectrum. For this case we denote the parameters by 
$\Delta_1,p_1,x_1$ and $a_1$. 
({\it iii}) The provider charges a price $p_2$ such that 
the delivered price is $\Delta_2=\Delta_0$ accounting for the
competition with the unlicensed band.
For this case, $x_2$ and $a_2$ are the number of customers 
in the licensed and unlicensed bands, respectively. 

We assume that $\Delta_1>\Delta_0=\Delta_2$ and will show that this implies $p_2x_2>p_1x_1$.  
This is a contradiction because the provider is assumed to maximize its revenue at $p_1$.
First observe that because $\Delta_1>\Delta_2$, $g(a_1)=\Delta_1$, 
and $g(a_2)=\Delta_2 $, it must be that $a_2<a_1$. 
Since $\Delta_1>\Delta_0$, the inverse demand curve at $x_0$ cannot be flat,
and so $x_2 + a_2 = x_0$.

Next, consider the function
$$
R(x)=(\Delta_2-l(x))x=(\Delta_0-l(x))x,
$$
which is the incumbent's revenue when the delivered price is fixed at $\Delta_2=\Delta_0$
as a function $x$.
Because $l(x)$ is a convex function, $R(x)$ is strictly concave. 
In case ({\it i}) (no unlicensed spectrum) the provider achieves 
the maximum revenue $R(x_0)$. Therefore, $R(x)$ is an increasing function 
from $0$ to $x_0$. Thus, since $x_2 \leq x_0$, it must be that 
$$
R(x_2)>R(x_2+a_2-a_1)=:R(x_3),
$$
where $x_3=x_2+a_2-a_1<x_2$. Thus, $x_3+a_1=x_2+a_2$.  
Now, because $\Delta_2=\Delta_0<\Delta_1$, we have
$$
x_2+a_2>x_1+a_1.
$$
Furthermore, because $x_3+a_1=x_2+a_2>x_1+a_1$, we have $x_3>x_1$.

Consider the difference 
$$R(x_3)-R(x_0)= R(x_3) -R(x_3 + a_1).$$
As seen on the left-hand side of Fig.~\ref{fig:123}, this is the difference 
between the areas $A$ and $B$, which is
\begin{equation}\label{eq:hc1}
R(x_3)-R(x_0)=x_3(l(x_3+a_1)-l(x_3)) - a_1(\Delta_0-l(x_3+a_1)).
\end{equation}

Similarly,  considering the difference $R(x_1) - R(x_1 + a_1)$ 
(given by the difference between  the areas of 
$A'$ and $B'$ on the right side of Figure~\ref{fig:123}), we have
\begin{equation}\label{eq:hc2}
\begin{split}
&p_1x_1-(\Delta_1-l(x_1+a_1))(x_1+a_1)\\
&=x_1(l(x_1+a_1)-l(x_1)) - a_1(\Delta_1-l(x_1+a_1)).
\end{split}
\end{equation}

Now since $x_3>x_1$ and $l(x)$ is convex, we obtain
$$
x_3(l(x_3+a_1)-l(x_3))> x_1(l(x_1+a_1)-l(x_1)).
$$
Furthermore, $\Delta_0<\Delta_1$ and $x_3+a_1>x_1+a_1$ implies
$$
a_1(\Delta_0-l(x_3+a_1))<a_1(\Delta_1-l(x_1+a_1)).
$$
Thus, the quantity in (\ref{eq:hc2}) must lower bound that in (\ref{eq:hc1}), giving
$$
R(x_3)-R(x_0)> p_1x_1-(\Delta_1-l(x_1+a_1))(x_1+a_1).
$$
Moreover, $(\Delta_1-l(x_1+a_1))(x_1+a_1)$ corresponds to the revenue the provider would achieve without the unlicensed band if it charges the price $\Delta_1-l(x_1+a_1)$.  Thus, $R(x_0)> (\Delta_1-l(x_1+a_1))(x_1+a_1)$. Therefore, 
$$
R(x_3)> p_1x_1,
$$
which is the desired contradiction to the fact that the provider optimizes its revenue. 
\qed

The next theorem states the drop in customer surplus with heterogeneous users.
\begin{theorem}\label{theo:heterogeneous}
Consider the heterogeneous model with a single incumbent and $P_h(0)\ge \lambda_h l(0)$. 
Suppose that without unlicensed spectrum both customer types are served at equilibrium.
Then there exists a $C_0>0$ such that when the bandwidth of unlicensed is increased 
from $C_0^{-}$ to $C_0^{+}$, the incumbent increases her price discontinuously
to serve high-type customers exclusively, which causes a drop in customer surplus.
\end{theorem}
{\bf Proof}: See Appendix~\ref{app:theorem3}.

Note that adding more unlicensed spectrum increases competition,
which decreases the revenue of the monopolist SP. 
Thus, this result shows that introducing additional spectrum 
as unlicensed band leads to a decrease in {\em both} 
the monopolist's revenue and consumer surplus.  

In Appendix~\ref{app:heterogeneous} we provide some numerical examples 
to illustrate how the consumer welfare depends on the unlicensed bandwidth. 
We show that there exists a range of parameters such that when $C_0^{+}$ bandwidth 
of unlicensed spectrum is introduced, customer surplus is strictly smaller 
than with only licensed spectrum. We next explain the intuition 
behind Theorem~\ref{theo:heterogeneous}.

Consider the model without unlicensed spectrum. The provider charges 
a price $p$ and by assumption serves both types of customers, i.e.,
all high-type customers and a fraction of the low-type customers. 
Let $R$ be the corresponding maximum total revenue. The provider could alternatively charge a higher price $p_H$ such that only high-type customers use the service.  Let the resulting revenue under $p_H$ be $R_H$, where $R_H<R$.

Now suppose an amount $C$ of unlicensed spectrum is introduced, 
which competes with the provider. As a result, the incumbent's maximized revenue  $R_C$ decreases in $C$. When $C$ is small the provider will 
also change the price $p$ by a small amount so that at this price 
both type of users are still served. When $C$ is large enough, however, the provider 
has an incentive to suddenly increase the price and in many cases could raise
$p$ up to $p_H$ to eliminate low-type customers and still obtain revenue $R_H$ 
from the high-type customers. When such a price increase occurs,  
the high-type customers pay a higher price and their surplus decreases. 
On the other hand, the low-type customers need to use the unlicensed band, 
which will be highly congested. As a result, their surplus decreases as well.


\section{Conclusions}\label{sec:conclude}

{\highlight 
We have studied a model for adding unlicensed spectrum to a market for wireless services 
in which incumbents have licensed spectrum. Our analysis has shown that there is
a range of scenarios in which adding unlicensed spectrum reduces overall welfare.
Specifically, this occurs when the additional unlicensed bandwidth is 
relatively small, so that it is readily congested.
In contrast, examples have shown that this decrease does not occur when
the additional bandwidth is assigned to a new or incumbent SP as licensed.
We emphasize that a key property of the unlicensed spectrum in our model is that
the congestion increases with the total traffic offloaded by {\em all} SPs,
which may be expected at low frequencies (TV bands). Hence our results
indicate that the success of applications in unlicensed bands at high frequencies may not translate to similar success at lower frequencies.

From a policy point-of-view these results indicate that if spectrum is abundant, 
then adding unlicensed spectrum should not lower welfare. Hence in settings
such as rural areas where demand is naturally limited, the type 
of congestion effects we consider may not be a significant issue. 
However, when spectrum is scarce, introducing relatively small amounts 
of white space might actually reduce welfare. Thus, in areas where demand is high, 
policies to limit congestion should be considered, such as establishing 
a market for a limited number of device permits as in \cite{Peha09}. 

These conclusions assume that the unlicensed spectrum offers services
that compete with the services provided in the licensed bands.
While that may create the most social welfare in some scenarios,
in other scenarios the unlicensed spectrum might be used to provide
other types of services (e.g., within a local area).
Indeed proponents of unlicensed spectrum have argued that
open access fosters innovation in technology and business models that may lead
to unforeseen uses of this spectrum (\cite{MilgromLevin}).
That possibility must then be weighed against the potential 
for innovation in licensed allocations in addition to the congestion effects
considered here.

}

\appendix
\section{Missing Proofs} \label{sec:appendix}

\subsection{\bf Proof of Theorem~\ref{thm:ss_wws}}\label{app:theo1}

In this proof we consider the incumbent as SP 1.  Before the unlicensed band is introduced, let $p_1^*$ be the  price charged by the incumbent and $x_1^* < 1$ the mass of customers served. After opening the unlicensed band with bandwidth $C$, let $x_1$ and $X^w$ be the number of customers using the licensed and unlicensed band respectively. Let $p_1$ be the price charged in the licensed band and
$P$ be the new delivered price. (See Figure~\ref{fig:addingwhitespace}).

Let $C_1$ be the value such that the corresponding congestion cost for $1-x_1^*$ customers is equal to $W$. That is
\begin{equation}\label{eq:c1}
g(1-x_1^*)=T_2+ \alpha_{C_1}(1-x_1^*)=W.
\end{equation}
\vspace{3mm}
\noindent
{\bf Proof of (i)}\\
\noindent
When $C \leq C_1$, it is straightforward to see that the unique equilibrium must be given by 
$p=p_1^*, x_1=x_1^*, X^w= g^{-1}(W)$, i.e.,  the unlicensed band does not  affect the 
price charged by the incumbent or the number of users it serves, and the overall welfare is constant.

Next we establish uniqueness of the equilibrium and its structure for  $C>C_1$.
First we prove that when $C>C_1$, at any equilibrium, all the customers will be served. To see this, assume that $x_1+X^w<1$. We then know that the delivered  price must be $W$, and so 
$$
g(X^w)= T_2+ \alpha_C{X^w}=W.
$$
Because $C>C_1$ we have $X^w>1-x_1^*$.  This shows that $x_1< 1-X^w<x_1^*$. Therefore,
$$p_1=W-l(x_1)>W-l(x_1^*)=p_1^*.$$
The incumbent, however, can charge a lower price to attract customers, who are currently unserved. Moreover,  total revenue is a concave function \footnote{ One can visualize the revenue of the incumbent $p_1^*x_1^*$ as the  area of the dashed rectangle on the left picture of Figure~\ref{fig:addingwhitespace}, where its lower-right corner runs on the line $l(x)$. It is straightforward to see that the revenue function is a concave function.}, and it is maximized at $p_1^*$. Thus by lowering $p_1$, which is greater than $p_1^*$, the incumbent can gain more revenue. This leads to a contradiction.

We now show that there is a unique equilibrium.  Assuming $C>C_1$, it follows that at any equilibrium
 $(p_1,x_1,X^w)$  must satisfy:
\begin{eqnarray}\label{eq:nash}
&& x_1+X^w=1 \nonumber \\
&& l(x_1)+p_1 = T_1+bx_1+p_1=P\le W\\
&& g(X^w)=T_2+ \alpha_CX^w=P\le W.\nonumber
\end{eqnarray}
From this one can derive a revenue maximization problem for the incumbent. Given $p_1$, $x_1(p_1)$ satisfying the above equations is a linear function of $p_1$. Thus, $\pi_1(p_1)=p_1x_1(p_1)$ is a quadratic  function of $p_1$ and therefore the incumbent's problem is
\begin{equation}\label{eq:maxrev}
\max_{p_1} \pi_1(p_1) \text{ subject to } p_1+ T_1+bx_1(p_1) \le W.
\end{equation}
This problem always has an unique  solution showing the uniqueness of the equilibrium.

\vspace{3mm}
\noindent
{\bf Proof of (ii)}\\
\noindent
In the remainder of the proof we derive the behavior of $S(C)$.  Observe that in  optimization problem (\ref{eq:maxrev}),
 depending on the parameters $T_1,T_2, b, W, C$, the solution must either be an interior point so that 
$\pi_1'(p_1)=0 $ or a boundary point satisfying $p_1+ T_1+bx_1(p_1)= W$.

Now consider the solution of the unconstrained  problem $\pi_1'(p_1)=0$. From~(\ref{eq:nash}), we have
$$
(b+\alpha_C)x_1+p_1=(T_2-T_1)+\alpha_C.
$$
Thus $\pi_1'(p_1)=(p_1\cdot x_1(p_1))'=0$ gives
\begin{equation}\label{eq:px_1}
p_1(C)=\frac{(T_2-T_1)+\alpha_C}{2};  x_1(C)= \frac{(T_2-T_1)+\alpha_C}{2(b+\alpha_C)}.
\end{equation}
Because $l(1)>g(0)$, we have $ T_2-T_1<b$, which  shows that
$$x_1=\frac{(T_2-T_1)+\alpha_C}{2(b+\alpha_C)}$$
 is  increasing in $\alpha_C.$
Therefore, $p_1+l(x_1)$ is increasing in $\alpha_C$. However, $\alpha_C$ is a decreasing function of $C$, thus
$p_1(C)+l(x_1(C)) 
$
is decreasing in $C$.
Furthermore, it is straightforward to see that when $C\rightarrow \infty$,
$$
p_1(\infty)+ l(x_1(\infty))= \frac{(T_2-T_1)}{2}+ T_1+ b \frac{(T_2-T_1)}{2b}=T_2<W,
$$
 and when $C\rightarrow 0$  $p_1(C)$  tends  to infinity because $\alpha_0=\infty$. Therefore, there exists an unique
$C^*$ such that $p_1(C^*)+l(x_1(C^*))=W$.

Now, if $C^* \le C_1$, then we define $C_2=\infty$, otherwise we define $C_2=C^*$. In both cases because
 $p_1(C)+l(x_1(C))$ decreases in $C$, we have  for all $C\in [C_1,C_2]$
$$
p_1(C)+l(x_1(C))>p_1(C_2)+l(x_1(C_2))=W.
$$
Therefore the unique equilibrium determined by~(\ref{eq:maxrev}) needs to  satisfy the condition  that the delivered price is $W$.

Now, when the delivered price is $W$,  observe that
$$
g(X^w)=W \Rightarrow X^w=C(W-T_2).
$$
Thus $X^w$ increases in $C$ and  $x_1=1-X^w$ decreases in $C$  and  by the same amount as $X^w$ increases. However, $l(x_1)<W$, which means that when $C$ increases the total mass of customers does not increase but some  customers switch  from a service with congestion cost $l(x_1)$ to a worse one (congestion cost of $W$)  and thus  the congestion cost increases and social welfare decreases.

\vspace{4mm}
Last, we  consider the case  $C>C_2$. We know that when $C>C_2$, the unique Nash equilibrium will satisfy $\pi_1'(p_1)=0$ and  we can use ~(\ref{eq:px_1}). In this case we know that the delivered price is $p_1+l(x_1)<W$, and all  customers are served. Therefore, social welfare is
$$
S(C)=p_1x_1+(W-p_1-l(x_1)), \text{ here } l(x_1)=T_1+bx_1.
$$
One can take the derivative of $S(C)$  with respect to $C$. Here, we simplify the formulation by a change of variables. Namely, let
$z=b+\alpha_C$ and $a=b+T_1-T_2 > 0$. We have $z'(C)=\alpha'(C) <0$ and
$$
p_1(C)=\frac{z-a}{2} ; x_1(C)=\frac{z-a}{2z}.
$$

A simple calculation yields
$$
S'(C)=z'(C)S'(z)=-\alpha'(C)\left( \frac{1}{4}+\frac{ab}{2z^2}+ \frac{a^2}{4z^2}\right).
$$
From this we see that $S'(C)>0$, therefore $S(C)$ is an increasing function.  This concludes the proof. \qed

\subsection{Extension of Theorem~\ref{thm:ss_wws}}\label{app:general}

Consider an environment with the boxed demand with consumer valuation $W$  as in Theorem~\ref{thm:ss_wws}, assume  an incumbent SP with licensed spectrum that does not  serve all of the demand. We assume $g(C,x)$ and $l(x)$ are strictly increasing and convex in $x$. Moreover, $g(C,x)$ is  continuously differentiable  and strictly decreasing in $C$ and $\lim_{C\rightarrow 0}g(C,x)=\infty$;   $\lim_{C\rightarrow \infty }g(C,x)<W$. Furthermore, to avoid a trivia scenario in which no consumers have incentive to use one of the two services, we also assume   $g(C,0)<W$ and $l(0)<W$.   We have the following result.
  
\begin{theorem}\label{theo:general}
If $x\cdot (g(C,1-x)-l(x))$ is concave in $x$, and the incumbent SP with licensed spectrum that does not  serve all of the demand, then 
\begin{itemize}
\item[(i)] For every $C\ge 0$ there is a unique equilibrium.
\item[(ii)] There exists $C_1<C_2$ such that as the amount of unlicensed spectrum $C$ increases from $C_1$ to $C_2$, the incumbent SP increases the price and social welfare decreases.
\end{itemize}
\end{theorem}

\noindent
{\bf Proof:} \\
First  notice that when $g(C,x)$ is increasing and linear in $x$, and $l(x)$ is an increasing convex function $x\cdot (g(C,1-x)-l(x))$ is concave and thus,  the result in Theorem~\ref{theo:general} holds.

The proof of Theorem~\ref{theo:general} is similar to that of Theorem~\ref{thm:ss_wws}. Namely, before the unlicensed band is introduced, let $p^*$ be the  price charged by the incumbent and  $x^* < 1$ the mass of customers served.  We have 
$$x^*=\argmax_x \;\;\;  x\cdot (W-l(x)).$$
Taking the derivative we obtain  
$
x^*(0-l'(x^*))+(W-l(x^*))=0.
$
This implies
\begin{equation}\label{eq:111}
W=l(x^*)+x^*\cdot l'(x^*)
\end{equation}
Let $C_1$ be the value such that the congestion cost in the unlicensed band for $1-x^*$ customers is equal to $W$. That is
\begin{equation}\label{eq:c11}
g(C_1,1-x^*)=W.
\end{equation}
Because of our assumption that $\lim_{C\rightarrow 0}g(C,x)=\infty$;   $\lim_{C\rightarrow \infty }g(C,x)<W$ and $g(C,x)$ is continuous in $C$, such $C_1$ exists.

Similar  to the proof of Theorem~\ref{thm:ss_wws},  when $C>C_1$ in a Nash equilibrium all customers are served in either the unlicensed or the licensed wireless service. Therefore, $x_1$ and $X^w$ are in an equilibrium if and only if.
\begin{eqnarray}
&& x_1+X^w=1 \label{eq:nash1} \\
&& l(x_1)+p_1 = g(C,X^w)=P\le W \label{eq:nash2}
\end{eqnarray}
The revenue of the incumbent SP (SP1) can be written as 
$$
Rev(x_1)=x_1\cdot p_1= x_1\cdot (g(C,1-x_1)-l(x_1)).
$$
Analogous to the proof of Theorem~\ref{thm:ss_wws}, because of the assumption that  $Rev(x_1)$ is  concave in $x_1$,  there is an unique equilibrium. Furthermore, let 
$$x_C=\argmax_x \;\;\; x\cdot (g(C,1-x)-l(x))$$
If  $g(C,1-x_C)\le W$, then at the equilibrium, number of consumer served by SP1 is $x_C$. If  on the other hand,  $g(C,1-x_C)>W$, then at the equilibrium $x_C$ cannot be at equilibrium because \eqref{eq:nash2} is violated. In this case at the equilibrium, the delivered price is $W$.

Now, consider  
$$x_{C_1}=\argmax_x \;\; x\cdot (g(C_1, 1-x)-l(x)).$$
Taking the derivatives, we have 
$$
x_{C_1}(-g'(C_1,1-x_{C_1})-l'(x_{C_1}))+(g(C_1, 1-x_{C_1})-l(x_{C_1}))=0.
$$
This is equivalent to 
$$
-g'(C_1,1-x_{C_1})\cdot x_{C_1}+ g(C_1, 1-x_{C_1})=l(x_{C_1})+x_{C_1}\cdot l'(x_{C_1}).
$$
Replacing $g(C_1, 1-x_{C_1})=W$ we get 
\begin{equation}\label{eq:222}
W=l(x_{C_1})+x_{C_1}\cdot l'(x_{C_1})+g'(C_1,1-x_{C_1})\cdot x_{C_1}
\end{equation}

Because $g'(C_1,1-x_{C_1})\cdot x_{C_1}>0$, thus, from \eqref{eq:111} and \eqref{eq:222} we have 
$$
l(x_{C_1})+x_{C_1}\cdot l'(x_{C_1})<l(x^*)+x^*\cdot l'(x^*)
$$
which implies $x_{C_1}<x^*$ because $l(x)$ is increasing and convex. Therefore 
$$g(C_1,1-x_{C_1})>g(C_1,1-x^*)=W.$$
Notice that because $x\cdot (g(C, 1-x)-l(x))$ is strictly concave and $g$ is continuously differentiable $C$, 
$x_{C}=\argmax  x\cdot (g(C, 1-x)-l(x))$ is continuous in $C$. Hence, $g(C,1-x_{C})$ is also continuous in $C$. Thus,  there exists $C_2>C_1$ such that 
$$
\text{for every  }C_1<C<C_2:   g(C,1-x_{C})>W.
$$
According to the argument above, for all $C_1<C<C_2$,  at the equilibrium, the delivered price  is~$W$.

Now,  let $x^*_C$ be the number of consumer at the equilibrium, we have 
$g(C,1-x^*_{C}) =W$ for $C_1<C<C_2$. Because $g(C,y)$ decreases in $C$, $ x^*_{C}$ also decreases as $C$ increases from $C_1$ to $C_2$. This means that as  $C$ increases from $C_1$ to $C_2$ SP will increase the price. As a result social welfare decreases.   This is what we need to prove.\qed

\subsection{Proof of Theorem~\ref{theo:heterogeneous}}\label{app:theorem3}

We first show that when the unlicensed spectrum bandwidth is large enough, the optimal revenue is obtained when the SP only serves high-type customers. Thus, because of the assumption that the SP serves a mixture of the two types in equilibrium before the unlicensed spectrum is introduced, there must be a  bandwidth of unlicensed spectrum, $C_0$, at which the SP switches from serving both types to only the  high type.

\begin{figure}[htbp]
\centering
\psfrag{Ph}{$P_h(x)$}
\psfrag{R}{$R_\delta$}
\psfrag{lam}{$\lambda_h\delta$}
\psfrag{delta}{$\delta$}
\psfrag{x}{$x_\delta$}
\psfrag{l}{$l(x)$}
\psfrag{lx}{$l(x_\delta)$}
\includegraphics[height=2.5in]{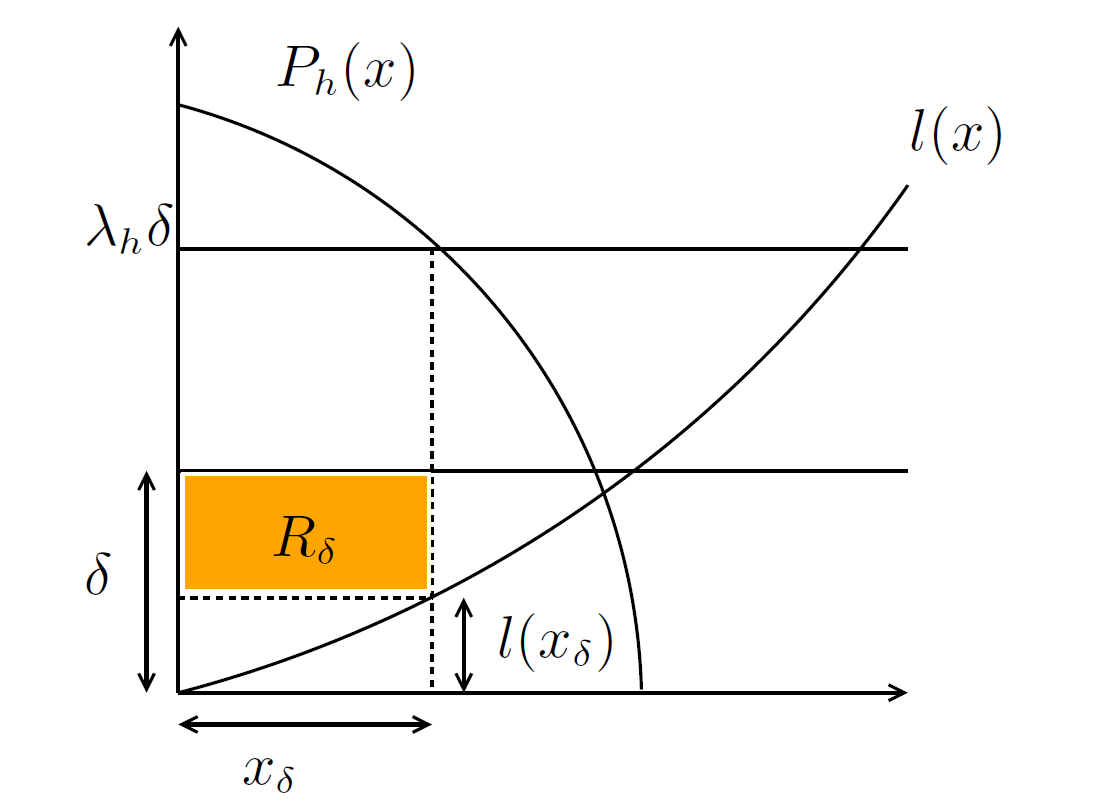}
\caption{Improving revenue by targeting only high typed customers}
\label{fig:delta}
\end{figure}

Given  $\delta>0$ let $x_\delta$ be the optimal point of
\begin{equation}
R_\delta = \max_x \{x\cdot(\delta-l(x))\}.
\label{eq:R_delta}
\end{equation}
$R_\delta$ is the shaded area in Figure~\ref{fig:delta}.

Note that $x_\delta$ is unique because $l(x)$ is convex. Furthermore, $x_\delta$ is a continuous function of $\delta$. 
When $\delta=l(0)$, $x_\delta=0$ and there exists  $\delta$ large enough such that $P_h(x_\delta) < \lambda_h\delta$.
Note that because $P_h(x)$ is  decreasing, $l(x)$ is  increasing  and  $\lambda_h l(0)< P_h(0)$, one can see that there exists $\delta^*$ such that 
$$
P_h(x_{\delta^*})= \lambda_h \delta^*.
$$

Now consider a situation where there is only unlicensed spectrum and there are only low type customers. Let $C_{\delta^*}$ be a value such that  if the bandwidth of unlicensed spectrum is $C_{\delta^*}$, then the the congestion cost of the unlicensed band is $\delta^*$.  We will show that in the setting with the incumbent SP when $C=C_{\delta^*}$ the optimal revenue of the incumbent is obtained by serving high type customers only.

To see this, observe that when serving both types of customers, the delivered price for low type customers cannot be higher than $\lambda_l \delta^*$. This is true because we know that when serving both types of customers, the users in the unlicensed spectrum are of the low type and because we have $C_{\delta^*}$ bandwidth of unlicensed spectrum,  the congestion in the unlicensed band is at most  $\delta^*$. Therefore, the optimal revenue that the incumbent SP can obtain while serving both types can be at most
$$
\max_x \{x\cdot\lambda_l(\delta^*-l(x))\}=\lambda_l R_{\delta^*},
$$
where $R_{\delta^*}$ is defined in (\ref{eq:R_delta}).

However, if the SP charges the price
$p=\lambda_h(\delta^*-l(x_{\delta^*}))$, then the equilibrium of the game is the following: no low type customers use the licensed spectrum and $x_{\delta^*}$ high customers use the service in the licensed band.  This is true because the congestion of the unlicensed band is $g(x)={\delta^*}$, the delivered price for low type customers in the licensed band is
$$
\lambda_l l(x_{\delta^*}) + \lambda_h(\delta^*-l(x_{\delta^*})) > \lambda_l l(x_{\delta^*}) + \lambda_l(\delta^*-l(x_{\delta^*})) =\lambda_l\delta^*.
$$
Thus, no low type customers would choose to use the licensed band.  On the other hand the delivered price for high type customers is  $\lambda_l{\delta^*}=P_h(x_{\delta^*})$. Therefore no high type customers would  use the unlicensed band either.

Now, in this case, the SP's revenue is
$$
p\cdot x_{\delta^*}= \lambda_h R_{\delta^*}> \lambda_l R_{\delta^*}.
$$

This shows that when $C=C_{\delta^*}$ the incumbent SP only serves high type customers. Therefore, there exists a $0<C_0<C_{\delta^*}$ such that if the unlicensed bandwidth is increased from $C_0^{-}$ to $C_0^{+}$,  the incumbent has an incentive to switch the class of customers and  target only the high class.

Consider such a transition. When $C=C_0^{+}$  the unlicensed band is open to all low type customers and they do not have other choices. Thus the delivered price for low type customers must be non-decreasing compared with when  $C=C_0^{-}$. Let $x_h, x_l$ be the number of customers of high and low types in the licensed band and $X^w$ be the number of customer in the unlicensed band at $C=C_0^{-}$. We have
$$
\lambda_l l(x_h+x_l) +p =  \lambda_l g(W^w).
$$
Thus, the delivered price for the high type customers at that time is
$$
\lambda_h l(x_h+x_l) +p <  \frac{\lambda_h}{\lambda_l}(\lambda_l l(x_h+x_l) +p) = \lambda_h g(W^w).
$$
This means that high type customers strictly prefer the licensed band to the unlicensed one.

Now, at $C=C_0^{+}$ the  quality of the  unlicensed band has worsened. Therefore, the incumbent  also has an incentive to raise the delivered price for high type customers.  This shows  that the delivered price for low type customers is non-decreasing and the delivered price for high type customers increases discontinuously. This concludes the  proof.   \qed

\section{Numerical Examples}\label{sub:numerical}
\subsection{\bf An example of social welfare loss} \label{app:example}
Consider the case where $T_1=T_2=0$, that is $l(x)=x, g(x)=\frac{x}{C}.$
That is $\alpha_C=\frac{1}{C}$. We will calculate $C_1, C_2, S(0)=S(C_1)$ and $S(C_2)$ as functions of $W$.

First we know that at the optimal monopoly price $p_1^*$, we have
$$
W-l(0)=2p_1^* \text{ and } W= x_1^* + p_1^*
$$
Thus  $x_1^*=p_1^*=W/2$, and according to~(\ref{eq:c1}), we have
$$
C_1=\frac{1-W/2}{W}.
$$
Note that because we assume that before unlicensed spectrum is introduced, the incumbent did not serve all customer, this can only happen when
$W<2$. Now,
$$
S(0)=S(C_1)=\frac{W^2}{4}.
$$
Next to calculate $C_2$ , we have
$$
p_1(C_2)+l(x_1(C_2))=\frac{1}{2C_2}+\frac{1}{2(C_2+1)} =W,
$$
which implies
$$
C_2=\frac{\sqrt{W^2+1}+1-W}{2W}>C_1.
$$
Thus,
$$S(C_2)=\frac{W^2}{2(\sqrt{W^2+1}+1)}.$$

For example if we consider $W=1$, then before  unlicensed spectrum is introduced, only half of the demand is met by a licensed spectrum with bandwidth 1.
Adding $C_2= \sqrt{2}/2\sim 0.7$ capacity of  unlicensed spectrum will create a new service that can serve  all the demand. However, because of the  congestion cost, the efficiency goes down to $\frac{S(C_2)}{S(C_1)}\sim 82.8\%$.

The worst example is when  $W=2$ then if $C_2=\frac{\sqrt{5}-1}{4}$ bandwidth of unlicensed spectrum is open then the efficiency can go down to
$\frac{S(C_2)}{S(C_1)}\sim 62\%$.


\subsection{A numerical example with multiple, symmetric incumbents}\label{multiple_incumbents}

We now give an example to show that
the conclusions from Theorem~\ref{thm:ss_wws} apply in more
general settings. Specifically, we consider a scenario in which there
is more than one incumbent SP. Additionally, we consider
a linear inverse demand given by $P(q)=1-\beta q$, where $\beta$
represents the elasticity of demand. Each SP has the same congestion cost
in her licensed spectrum, with $l_i(x)=l(x) = x$ for all
$i\in\mathcal{N}$. The congestion cost in the unlicensed band is given
by $g(x) = \frac{x}{C}$.

\begin{figure}[htbp]
\centering
\includegraphics[ height=2.5 in]{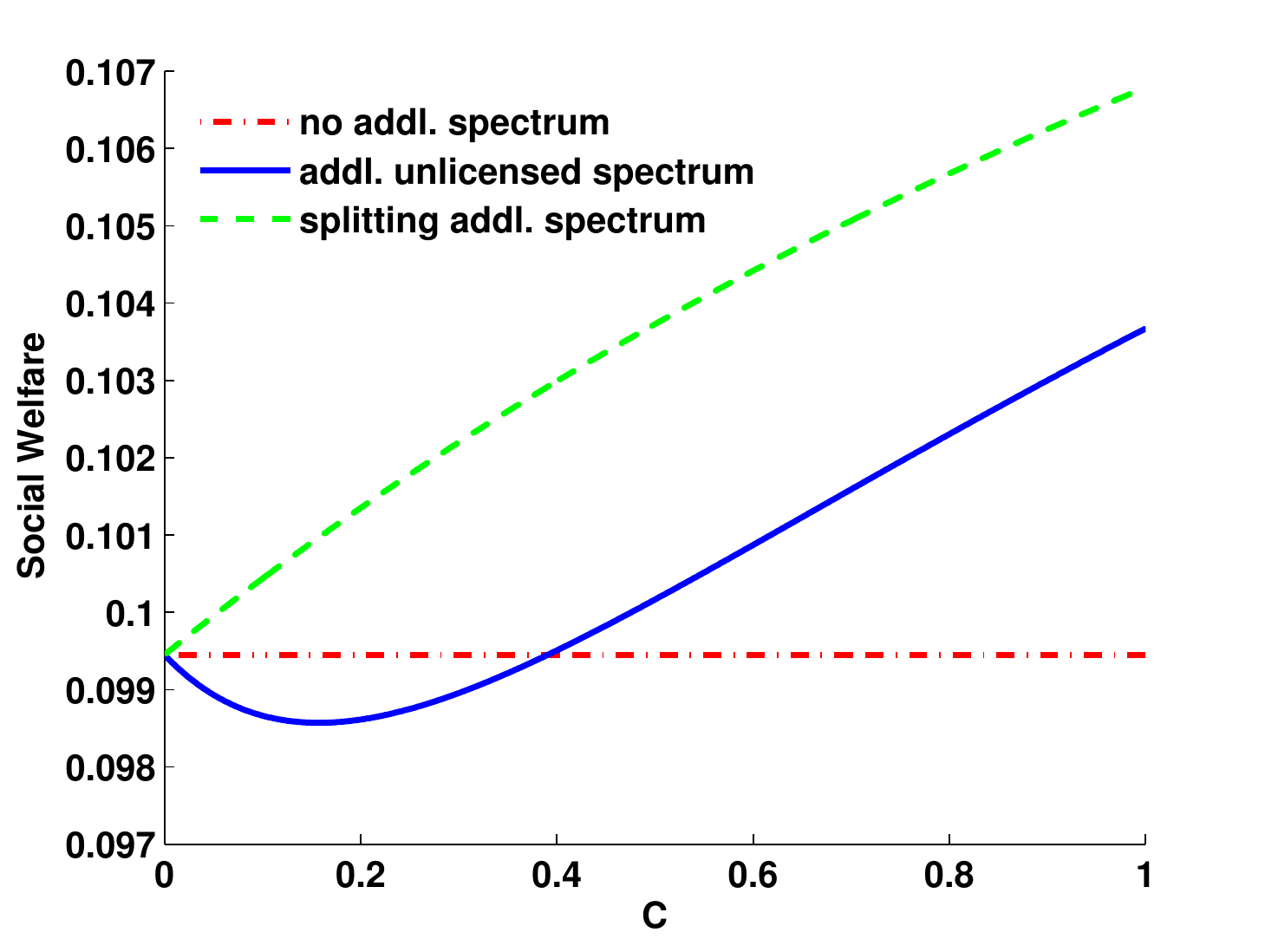}
\caption{The social welfare in different scenarios
as a function of additional capacity $C$
in a symmetric linear network with $N=2$ and $\beta=4$.} \label{fig:lin_sym}
\end{figure}

For such a model there is a unique symmetric equilibrium. One can actually explicitly write down the social welfare with there are $N$ SPs either with or
without an additional band of unlicensed spectrum.  
Here, we provide a numerical example showing  the social welfare  decreases when additional capacity is added, i.e., Braess's paradox occurs.
This example is shown in Fig.~\ref{fig:lin_sym}, where
the solid curve is the welfare with additional unlicensed spectrum as
a function of the amount of additional spectrum.

We can also determine the social welfare for a scenario where instead of making the $C$ units of capacity freely available, we divide this capacity evenly among the existing $N$ SPs.   We model this by again assuming that $l(x)$ is given by the customer mass  per unit capacity for each licensed band, where initially the capacity is normalized to one.
Hence, after giving each SP $C/N$ additional units of capacity, the
 new congestion function is $\tilde{l}(x)=\frac{1}{1+C/N}x$. This
quantity is also shown in Fig.~\ref{fig:lin_sym}. In this case
dividing up the spectrum in this manner improves the welfare for all
values of $C$. This suggests that in cases where Braess's paradox
occurs, licensing the spectrum to existing SPs can be socially
more efficient.


\subsection{Numerical results for the heterogeneous model}\label{app:heterogeneous}

\begin{figure}[htbp]
\centering
\includegraphics[height=4in]{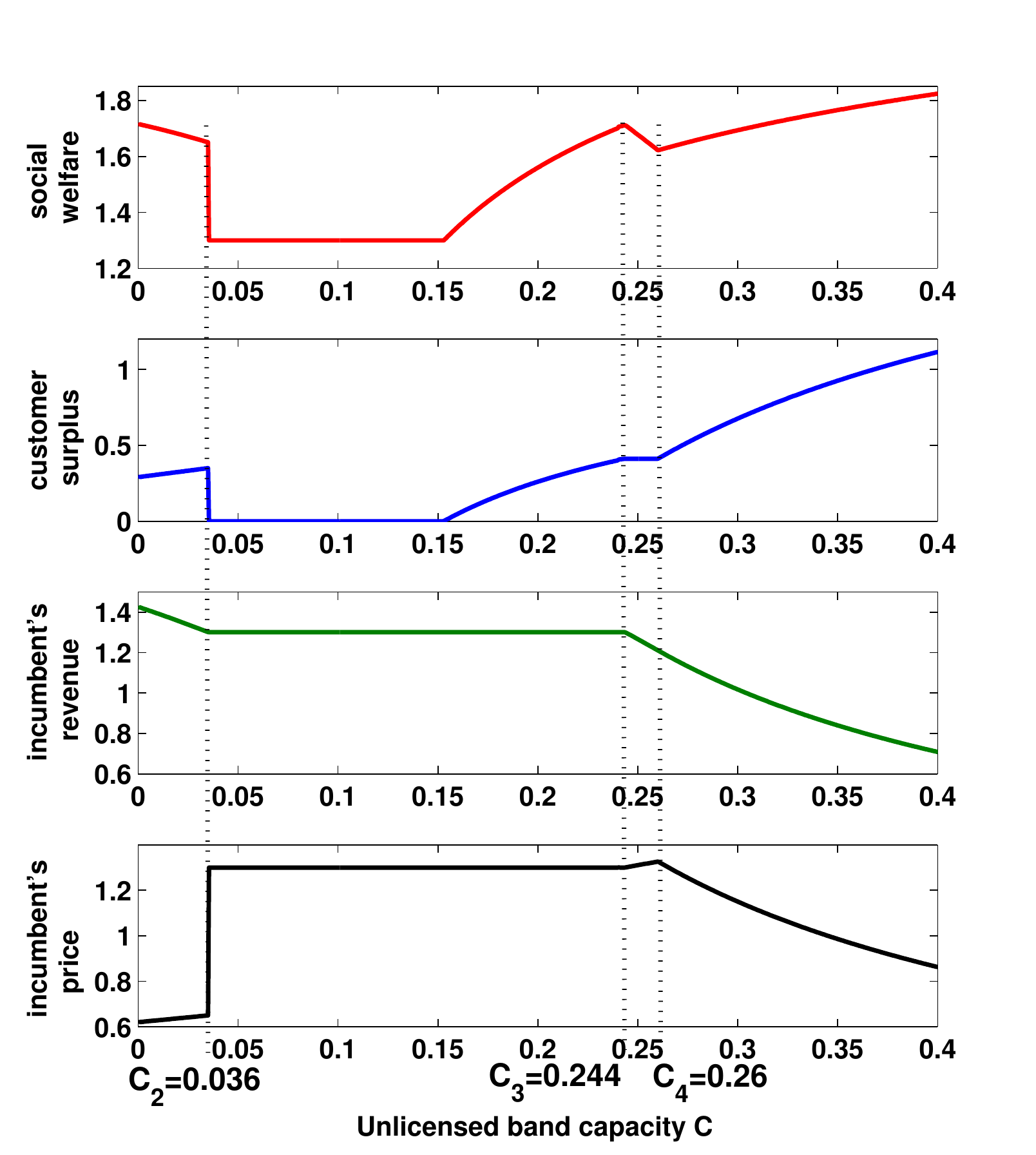}
\caption{An example where the incumbent served both classes initially}
\label{fig:eg_heter}
\end{figure}

We now show a numerical example illustrating the social welfare in a heterogeneous model. The specific numerical values are the following.
Let $W_h=1.6$, $Q_h=1$, $W_l=0.85$ and $Q_l=1.3$,  $\lambda_h=0.4$ and  $\lambda_l=0.1$.
Set $l(x)=x$ and $g(x)=x/C$.  Without unlicensed band, it can be shown that the incumbent SP would set price $p_0=0.62$ to serve all of low class customers to maximize her revenue. The numerical results of social welfare, customer surplus and incumbent's price and revenue are plotted against the unlicensed band capacity $C$ in Fig. \ref{fig:eg_heter}.

In this numerical example, we found the social welfare, as a function of the capacity $C$ in the unlicensed band, is not monotone. As shown in Fig. \ref{fig:eg_heter}, there are two regions of $C$, $[0,C_2]$ and $[C_3,C_4]$ where social welfare decreases.  Note that $C_2'>C_2$ for the parameters in this example, so we define $C_2':=C_2$.

Comparing social welfare and incumbent's price in Fig. \ref{fig:eg_heter}, we find that the two regions where social welfare decreases are also where the incumbent's price rises. This phenomenon is reminiscent to that in the model with homogeneous customers and can be explained in a similar way. Namely, the incumbent may benefit from raising her price since this may reduce the congestion in her licensed band and worsen the quality of service in the unlicensed band.

In particular, there are three stages as $C$ increases. Facing the competition from the service in the unlicensed band as $C$ increases, the incumbent will eventually ``retreat" from serving low class and suddenly increases its price to serve high class customers only to gain higher revenue.  This corresponds to the jump in the monopoly's price and the drop in social welfare and customer surplus in Fig.~\ref{fig:eg_heter} at $C_2$.

The first stage corresponds to $C\in[0,C_2]$. In this stage, the service in the unlicensed band and that of the incumbent's licensed band will be competing on \emph{low} class customers while the incumbent still serves all of the high class customers. Here we have the same observation that social welfare decreases as a result of the rise of the monopoly's price and congestion in the unlicensed band.

The second stage is when $C\in[C_2,C_3]$. This is the stage in which the market is sorted. Namely, the unlicensed band serves only low class and licensed band serves high class customers. Thus increasing capacity $C$ has no impact on high class customers but improving the congestion in the unlicensed band. Therefore, the social welfare is constant or increasing in this stage.

Finally, when $C\in[C_3, \infty]$, the unlicensed band and licensed band will be competing on the \emph{high} class customers while all of the low customers are being served in the unlicensed band. Similarly,  the observation is that the social welfare decreases first as the increase of the monopoly's price until $C$ reaches $C_4$  and then eventually starts to increase as the quality of service in the unlicensed band improves.

\bibliographystyle{apalike}
\bibliography{ref}

\end{document}